\documentclass{article}

\usepackage{arxiv}

\usepackage[utf8]{inputenc} 
\usepackage[T1]{fontenc}    
\usepackage{hyperref}       
\usepackage{url}            
\usepackage{booktabs}       
\usepackage{amsfonts}       
\usepackage{nicefrac}       
\usepackage{microtype}      
\usepackage{lipsum}		
\usepackage[final]{graphicx}
\usepackage[numbers]{natbib}
\usepackage{doi}
\usepackage{multirow}
\usepackage{float}
\usepackage{subcaption}
\usepackage{placeins}
\usepackage{amsmath}
\usepackage{pifont}
\usepackage[table]{xcolor}

\usepackage{array}
\newcolumntype{L}[1]{>{\raggedright\let\newline\\\arraybackslash\hspace{0pt}}m{#1}}
\newcolumntype{C}[1]{>{\centering\let\newline\\\arraybackslash\hspace{0pt}}m{#1}}
\newcolumntype{R}[1]{>{\raggedleft\let\newline\\\arraybackslash\hspace{0pt}}m{#1}}

\title{Differentiable Black-box and Gray-box Modeling of Nonlinear Audio Effects}

\author{
    Marco Comunità\\
    \texttt{m.comunita@qmul.ac.uk}
    \AND Christian J. Steinmetz \And Joshua D. Reiss
    \AND \\
    Centre for Digital Music\\
    Queen Mary University of London, UK\\
}

\date{}

\renewcommand{\shorttitle}{\textit{arXiv} Template}

\hypersetup{
pdftitle={A template for the arxiv style},
pdfsubject={q-bio.NC, q-bio.QM},
pdfauthor={David S.~Hippocampus, Elias D.~Striatum},
pdfkeywords={First keyword, Second keyword, More},
}

\begin{document}
\maketitle

\begin{abstract}
Audio effects are extensively used at every stage of audio and music content creation.
The majority of differentiable audio effects modeling approaches fall into the black-box or gray-box paradigms; and most models have been proposed and applied to nonlinear effects like guitar amplifiers, overdrive, distortion, fuzz and compressor. 
Although a plethora of architectures have been introduced for the task at hand there is still lack of understanding on the state of the art, since most publications experiment with one type of nonlinear audio effect and a very small number of devices. 

In this work we aim to shed light on the audio effects modeling landscape by comparing black-box and gray-box architectures on a large number of nonlinear audio effects, identifying the most suitable for a wide range of devices. 
In the process, we also: introduce time-varying gray-box models and propose models for compressor, distortion and fuzz, publish a large dataset for audio effects research - ToneTwist AFx\footnote{\url{https://github.com/mcomunita/tonetwist-afx-dataset}} - that is also the first open to community contributions, evaluate models on a variety of metrics and conduct extensive subjective evaluation. Code\footnote{\url{https://github.com/mcomunita/nablafx}} and supplementary material\footnote{\url{https://github.com/mcomunita/nnlinafx-supp-material}} are also available.
\end{abstract}


\section{Introduction}
Audio effects are central for engineers and musicians to shape the timbre, dynamics, and spatialisation of sound~\cite{wilmering2020history}.
For some instruments (e.g., electric guitar) the processing chain is part of the artist’s creative expression \cite{case2010recording} 
and music genres and styles are also defined by the type of effects adopted 
\cite{blair2015southern, williams2012tubby}
; with renowned musicians adopting specific combinations to achieve a unique sound \cite{prown2003gear}.
Therefore, research related to audio effects, especially with the success of deep learning 
\citep{bengio2017deep} 
and differentiable digital signal processing (DDSP) \citep{engel2020ddsp}, is a very active field \citep{comunita19afxresearch}, with many avenues of investigation open, like:
audio effects classification and identification 
\citep{
comunita2021guitar, 
hinrichs2022classification}, 
settings or coefficients estimation and extraction 
\citep{
kuznetsov2020differentiable, 
nercessian2020neural, 
colonel2022direct}, 
modeling 
\citep{
martinez2021deep, 
wright2023adversarial}, 
processing graph estimation 
\citep{
lee2023blind, 
lee2024searching} 
and style transfer 
\citep{
steinmetz2022style, 
vanka2024diff, 
steinmetz2024st}, 
automatic mixing 
\citep{
steinmetz2021automatic, 
martinez2022automatic}
or reverse engineering of a mix 
\citep{
colonel2022reverse, 
colone2023reverse}. 

One of the most active, driven by interest in emulating vintage or costly analog devices, is audio effects modeling (virtual analog), often categorized into white-, gray-, and black-box approaches (see Sec.~\ref{sec:background});
and, when using techniques that support gradient backpropagation, referred to as differentiable white-, gray- and black-box.
Differentiable methods, leveraging data for gradient descent, are widely used to learn audio effect models. 
Efforts focus on developing architectures that are both accurate and computationally efficient while being expressive enough to model diverse effect categories and devices.

Of the three macro categories of audio effects - linear (e.g., EQ, reverb), nonlinear (e.g., distortion, compression), and modulation (e.g., phaser, flanger) - nonlinear effects have received the most attention, with architectures mainly proposed for black- and gray-box methods (see Secs.~\ref{sec:back-diff-black-box} and \ref{sec:back-diff-gray-box}).
Most nonlinear audio effects models focus on a single category (e.g., guitar amplifiers, distortion, compressors) and are evaluated on a single or limited set of devices.
Despite the effort, there is no clear state of the art yet that identifies the best-performing approach, particularly for developing architectures that can reliably model all types of nonlinear effects across a wide range of devices.
Also, all previous works train and evaluate on data recorded from the same sources (e.g. a specific electric guitar or bass) and very similar content, playing techniques and style, not allowing to reach a reliable conclusion about the generalization capabilities of a trained model.
In most cases, training and testing data is normalized without verifying the input dynamic range, not ensuring trained models to be accurate across signal levels.

Therefore, to overcome these limitations, assess the state of the art, and identify research directions, we conduct extensive experiments.
In doing so, our work contributes mainly in the following ways:
\begin{itemize}
    \item Evaluate and compare differentiable black-box and gray-box approaches for modeling nonlinear audio effects like guitar amplifiers, overdrive, distortion, fuzz, and compression (Sec.~\ref{sec:experiments}).
    \item Introduce time-varying gray-box models and propose models for compressor/limiter, distortion and fuzz effects (Sec.~\ref{sec:gb-models}).
    \item Publish a large dataset of dry-input/wet-output pairs for audio effects research that is also the first open to community contributions (Sec.~\ref{sec:tonetwist}).
    \item Evaluate the models on a variety of metrics (Sec.~\ref{sec:obj_eval}) and conduct extensive subjective evaluation (Sec.~\ref{sec:subj_eval}).
\end{itemize}

The remainder of the paper is organized as follows. 
In Section~\ref{sec:background}, we outline non-differentiable and differentiable white-, gray-, and black-box approaches to audio effects modeling.
Sections~\ref{sec:bb-models} and \ref{sec:gb-models} describe the black- and gray-box architectures included in our experiments, while the ToneTwist AFx dataset is described in \ref{sec:tonetwist}. 
Experiments and results are illustrated and discussed in Sections \ref{sec:experiments} and \ref{sec:results}, and objective and subjective evaluations are presented in \ref{sec:eval}. 
We conclude the manuscript in \ref{sec:conclusion}.

\section{Background}
\label{sec:background}
\subsection{Nonlinear Audio Effects Modeling}
\label{sec:back-nnlinafx-modeling}


The first thorough investigation of nonlinear distortion for musical applications is Le Brun's 1979 work on waveshaping \citep{le1979digital} and Roads' subsequent tutorial \citep{roads1979tutorial}.
In the following decades, numerous techniques were developed to digitally simulate or emulate nonlinear analog circuits and devices.
This is why we now have approaches based on: 
solving or approximating the equations describing a system's behavior, 
wave digital filters, 
state-space modeling, 
Volterra series 
or Wiener-Hammerstein models, 
just to name a few.
And, with the advancement of machine learning, neural networks based modeling 
and differentiable DSP 
have been added to the state of the art.

Audio effects modeling approaches can be categorized into white-, gray-, and black-box, based on the degree of prior knowledge applied, and further classified as differentiable if they support gradient backpropagation.
The following sections review previous work on nonlinear audio effects modeling and describe emulation paradigms and techniques.

\subsection{White-box Modeling}
\label{sec:back-white-box}

White-box modeling relies on full system knowledge, using differential equations to describe its behavior and numerical methods to solve them in the continuous or discrete domain. 
These models provide accurate results, making them ideal for replicating specific analog devices.
Although, they can be time-consuming to develop, require exact knowledge of the equations describing nonlinear elements and access to the circuit schematic, and can result in substantial computational load.

Simple systems can be modeled manually by solving the equations 
\citep{
yeh2007simulation, 
d2014generalized, 
esqueda2017virtuallockhart}; 
but, for more complex cases, there exist general-purpose frameworks like: 
state-space models 
\citep{mavcak2012real, holters2015generalized, yeh2009automated}, 
wave digital filters 
\citep{
werner2015wave, 
de2009virtual, 
d2012new}, 
port-hamiltonian systems 
\citep{falaize2016passive}.

In practice, solving \textbf{equations} have mostly been applied to a small subset of single nonlinear processing blocks like: 
vacuum tube 
\citep{
sapp1999simulation, 
karjalainen2006virtual, 
dempwolf2011physically} 
or transistor stages 
\citep{d2019fast}, 
diode clippers 
\citep{
yeh2007simulation, 
germain2015design, 
d2019fast}, 
transformers 
\citep{macak2011nonlinear}, 
moog ladder filter 
\citep{
d2013improved, 
d2014generalizedI, 
d2014generalizedII} 
or the Buchla 259 wavefolder 
\citep{esqueda2017virtualbuchla}.
But there are also examples of literature modeling complete devices with this method 
\citep{
yeh2007simplified, 
macak2013guitar}. 

The most studied white-box modeling framework is \textbf{wave digital filters} (WDF), which emulates analog circuits using traveling wave (scattering) theory and wave variables (instead of voltages and currents), preserving properties like passivity and energy conservation for stable, robust digital implementation.
\footnote{\url{https://ccrma.stanford.edu/~dtyeh/papers/wdftutorial.pdf}}
\citep{fettweis1986wave}.

Beside single nonlinear blocks like: 
vacuum tube 
\citep{
karjalainen2006wave, 
yeh2008simulating, 
d2012wave}
or transistor stages 
\citep{
yeh2008simulating, 
yeh2009digital, 
bogason2018modeling}, 
diode clippers 
\citep{
yeh2008numerical, 
yeh2008simulating, 
werner2015improved, 
bernardini2017biparametric}, 
opamps 
\citep{
yeh2009digital, 
bogason2017modeling}, 
nonlinear filters 
\citep{
werner2015resolving, 
bogason2018modeling} 
and transformers 
\citep{
pakarinen2009wave}; 
we also find examples of WDFs applied to complete devices like: 
overdrive and distortion pedals 
\citep{
yeh2009digital, 
bernardini2020wave},
amplifiers 
\citep{
pakarinen2009wave, 
de2009virtual}, 
limiters 
\citep{raffensperger2012toward}.

Similar observations are valid for \textbf{state-space methods}, which represent an electronic circuit as a system of first-order differential equations, describing the circuit in terms of state variables, inputs, and outputs. 
Using matrix algebra to model components relationships, this method enables efficient digital simulation of complex analog systems.
Methods developed for the simulation of state-space systems include the K-method 
\citep{
yeh2006discretization, 
yeh2009digital}
the Nodal K-method (or NK-method) and the Discrete K-method (or DK-method or Nodal DK-method) 
\citep{
yeh2009digital, 
yeh2009automated}. 

Examples of state-space methods applied to single nonlinear blocks include: 
vacuum tubes, 
transistors, 
diode clippers 
and opamps
\citep{
yeh2008simulating, 
yeh2009digital, 
yeh2009automated}, 
or transformers 
\citep{holters2016circuit}.
But also find examples of emulation of complete devices like: 
overdrive and distortion 
\citep{
yeh2009digital, 
holmes2015improving}, 
fuzz 
\citep{
holmes2017comparison, 
bennett2022state},
compression 
\citep{kroning2011analysis},
or amplifiers 
\citep{
cohen2010real, 
macak2012simulation}.

\textbf{Port-Hamiltonian} approaches emulate analog circuits using Hamiltonian mechanics, focusing on energy conservation \citep{van2006port}.
The only application of this approach to nonlinear audio effects is limited to single transistor stages and diode clippers.
\citep{falaize2016passive}.\\

\subsection{Black-box Modeling}
\label{sec:back-black-box}

Black-box approaches require minimal system knowledge, and mostly rely on input-output data; simplifying the process to collecting adequate data.
However, such models often lack interpretability and might entail time-consuming optimizations.
The main non-differentiable black-box methods are based on Volterra series and dynamic convolution.

\textbf{Volterra series} represent a nonlinear system as an infinite series of integral operators, analogous to a Taylor series expansion but in the context of systems and signals. 
The output is given by a sum of convolutions of the input signal with a series of kernels (functions), and the modeling procedure mainly involves extracting such kernels from input-output measurements.
Although it has been applied to a range of nonlinear blocks 
\citep{
helie2006use, 
yeh2008numerical} 
and effects 
\citep{
carini2015legendre, 
tronchin2013emulation, 
schmitz2017hammerstein, 
orcioni2017multivariance, 
carini2019nonlinear}
, this method is sufficiently accurate only for weakly nonlinear systems, making it not applicable to most nonlinear audio effects.

The same applies to \textbf{dynamic convolution}
\citep{
kemp1999analysis, 
primavera2012approximation}
, which adjusts impulse responses or processing kernels based the present/past input amplitude to model nonlinear or hysteretic behavior.

\subsection{Gray-box Modeling}
\label{sec:back-gray-box}

Gray-box approaches combine a partial theoretical structure with data - typically input/output measurements - to complete the model.
They reduce prior knowledge requirements while maintaining interpretability through a block-oriented approach; although, the specific structure, measurements and optimization procedures, are critical to achieve a good approximation.

For nonlinear effects like amplifiers and distortion, models are typically defined as an interconnection of linear filters and static nonlinearities, such as: Hammerstein models (static nonlinearity followed by linear filter), Wiener models (linear filter followed by static nonlinearity) or Wiener-Hammerstein models (static nonlinearity in between two linear filters).
Although, for greater accuracy \textbf{Wiener-Hammerstein models} have often been extended to include: 
non-static nonlinearities (i.e., hysteresis and memory) 
\citep{
eichas2015block, 
eichas2016black, 
eichas2018gray},
or a cascade of pre- and power-amp models 
\citep{
kemper2014musical, 
eichas2017block}.
There exist more complex arrangements made of cascaded and parallel blocks (see \citep{schoukens2019nonlinear} for more examples) and, for nonlinear effects, 
parallel polynomial 
\citep{
novak2009nonlinear, 
cauduro2012reduced} 
and Chebyshev 
\citep{
novak2010chebyshev, 
bank2011computationally} 
Hammerstein models have also been adopted.

To model dynamic range compression, all proposed methods 
\citep{
ramos2011block, 
eichas2016modeling} 
are based on the block-oriented architecture described in 
\citep{zolzer2002dafx} 
and analyzed in details in 
\citep{giannoulis2012digital}.

\subsection{Differentiable White-box Modeling}
\label{sec:back-diff-white-box}

With the recent rise of machine learning the research community started applying neural networks and DDSP to white-, gray- and black-box modeling.
In \citep{parker2019modelling}, the authors use a multi-layer perceptron (MLP) as a state-space model, called a state trajectory network, which predicts the output based on the input signal and an internal state.
Since state-space models are based on circuit variables, when applying the method to a first-order and a second-order diode clipper the authors adopt input and internal voltages as training signals to predict output voltage.
This approach is then extended in \citep{peussa2021exposure} using a gated recurrent unit (GRU).\\
In \citep{esqueda2021differentiable}, the authors introduce differentiable white-box virtual analog modeling, using backpropagation to optimize component values in analog circuits, applied to find optimal resistor and capacitor values that approximate the frequency response of an RC filter and tone-stack.\\
The idea of learning ordinary differential equations using differentiable methods is applied in \cite{wilczek2022virtual} to learn the governing equations of diode clippers.\\
In \citep{parker2022physical}, the authors use recurrent neural networks with fast convolutional layers to model partial-differential equation governed systems, applying it to lossy dispersive string, 2D wave equation, and tension-modulated string.\\
Analogously to \citep{parker2019modelling} for state-space models, \citep{chowdhury2022emulating} introduces differentiable wave digital filters, using an MLP to learn the input-output relationship of wave variables in a diode clipper.

\subsection{Differentiable Black-box Modeling}
\label{sec:back-diff-black-box}

Black-box modeling is the most common approach for differentiable audio effects, and it has been applied to most linear, non-linear, and modulation effects, although with varying success.

The first example of neural networks adoption for a modeling task is \citep{mendoza2005emulating}, where the author uses an \textbf{MLP} to model overdrive.
Following this, recurrent networks were proposed for guitar amplifiers modeling, with \textbf{echo state machines} \citep{holzmann2009reservoir} and \textbf{NARX} networks \citep{covert2013vacuum} being adopted due to their easier and more stable training with respect to early LSTMs.
Although, once the training issues were overcome, recurrent networks like \textbf{LSTMs} \citep{zhang2018vacuum, schmitz2018nonlinear} have become a common approach, with different nonlinear effects, architectures and conditioning methods being further studied in 
\citep{wright2019real, wright2020real, wright2020perceptual, chowdhury2020comparison, simionato2022deep, simionato2024comparative, yeh2024hyper}.

In \citep{schmitz2019objective}, an \textbf{hybrid CNN and LSTM} architecture is explored for guitar amps emulation, while \citep{ramirez2019modeling} is the first example of \textbf{Encoder/Decoder} (ED) architecture applied to nonlinear effects, although limited by a low sampling rate (16~kHz) and non-causal implementation.
Improvements on this ED architecture were further explored in \citep{ramirez2019general, martinez2020deep} and applied to dynamic range compression among other effects.

\textbf{CNN} architectures took longer to be explored as a modeling approach. 
In the first examples 
\citep{damskagg2019real, damskagg2019deep} 
the authors used a non-autoregressive WaveNet 
\citep{rethage2018wavenet}, also known as a gated convolutional network (GCN), which is a special case of temporal convolutional network (TCN) \citep{bai2018empirical} that uses gated activations. 
These works focused on guitar amplifiers, overdrive, and distortion effects, and achieved results equal or better than recurrent networks in a subjective test \citep{wright2020real}.
TCNs have been proposed for compressor modeling in \citep{steinmetz2022efficient}, while GCNs have been extended with temporal feature-wise linear modulation (TFiLM) \citep{birnbaum2019temporal} in \citep{comunita2023modelling} to model time-varying nonlinear effects like fuzz and compressor, and shown in both cases to outperform recurrent networks and standard GCNs.

ED-based architectures have also been further explored in the time \citep{simionato2023fully} and spectrogram \citep{hawley2019profiling, mitchell2020exploring} domain for compressor modeling.
More recently, with the success of \textbf{state-space models} 
\citep{gu2021combining, gu2021efficiently, gu2023mamba} 
in general time series modeling tasks, the application to effects modeling like compressor and overdrive were also explored 
\citep{yin2024modeling, simionato2024comparative, simionato2024modeling}.
It is also worth noticing how LSTMs and GCNs have also been adopted for patents on differentiable nonlinear audio effects modeling for commercial products \citep{borquez2022neural} (Neural DSP Technologies).

\subsection{Differentiable Gray-box Modeling}
\label{sec:back-diff-gray-box}

With the development of DDSP \citep{hayes2024review}, it has become possible and also desirable to explore differentiable gray-box approaches to audio effects modeling. 
The main advantages being the reduced number of parameters and higher computational efficiency w.r.t. neural networks as well as the higher degree of interpretability.

Few studies explore block-based differentiable modeling, with most extending W-H models for nonlinear effects (guitar amps, overdrive, distortion, fuzz) 
\citep{
kuznetsov2020differentiable, 
nercessian2021lightweight, 
colonel2022reverse, 
miklanek2023neural, 
yeh2024ddsp} 
or implementing a differentiable version
\citep{
lee2024searching, 
yu2024differentiable, 
colonel2022approximating, 
wright2022grey} 
of previously suggested implementations of feed-forward digital compressors 
\cite{zolzer2002dafx, giannoulis2012digital}.

The first \textbf{W-H model}-inspired approach \citep{matsunaga2018digital} tests overdrive emulation, with input and output convolutional layers used to learn FIR filters and LSTM layers used to learn a nonlinearity.
A similar approach \citep{taylor2020latent} implements guitar amps emulation as a series of six learnable Wiener models (i.e., linear filters followed by nonlinearity). 
Here the filters are implemented as fixed band-pass, while the scaling factor for each band component is learned, and the nonlinear block is set to a soft-sign. 
Tested on several amplifiers, it allows to implement \textit{hybrid models} by interpolating coefficients from different amps.

\textbf{DDSP} for gray-box modeling is first explored in \citep{kuznetsov2020differentiable}, where differentiable linear time-invariant filters and a nonlinearity form a W-H model that emulates the Boss DS-1 distortion pedal.
LTI filters are implemented either as FIRs with 128 taps or second-order IIRs (biquads), while an MLP is used as a memoryless nonlinearity.\\
For their model of the Boss MT-2 distortion pedal, Nercessian et al. \citep{nercessian2021lightweight} adopt a cascade of 40 hyperconditioned filtering stages, each made of biquad, gain and $\tanh$ nonlinearity.
Hyperconditioning \citep{ha2022hypernetworks, serra2019blow} involves using neural networks to adjust the internal coefficients or weights of a differentiable block, such as a neural network or DDSP block. 
In this case, a parametric model is obtained by conditioning, through affine transformations, biquad coefficients and gain.
Therefore, the controllers we use in our work (see Sec.\ref{sec:gb-dist}) are a form of hyperconditioning. 
Though interpretable, the long sequence of nonlinear filters doesn't closely match the actual circuit or is easily observable.\\
Also Colonel et al. \citep{colonel2022reverse} explore differentiable W-H models with a focus on proposing several parametric nonlinearities: harmonic sum of $\tanh$ functions, power sum of $\tanh$ functions, Fourier series and Legendre polynomials.
LTI filters are implemented as differentiable 20-band FIR graphic EQs. 
While good results are achieved, the model is limited to single input-output pairs (overfitting to one example) and is non-parametric.\\
In \citep{miklanek2023neural}, Miklanek et al. proposed a hybrid differentiable guitar amplifier model, combining black-box and gray-box approaches. The pre- and power amplifiers are modeled using recurrent and fully connected layers, while a differentiable parametric tone-stack model, with a known processing function, is placed between them.\\
A recent guitar amplifier model in \citep{yeh2024ddsp} features a detailed, differentiable parametric processor for each stage of a typical amplifier's analog circuit. 
This includes several W-H stages for the pre-amplifier, a tone-stack \citep{miklanek2023neural}, a custom AB push-pull power amplifier model, and a custom output transformer model. 
Despite its advanced design, the best-performing gray-box arrangement significantly underperforms compared to a single-layer recurrent network, with more trainable parameters (10.1k vs. 8k), resembling a black-box model in size.

\subsection{State of the art}
\label{sec:back-sota}

As we have seen in the previous sections there is a vast literature on nonlinear audio effects modeling detailing a large number of approaches and paradigms, which include both differentiable and non-differentiable implementations.
Generally speaking, the vast majority of the effort for non-differentiable methods is focused on white-box models, where the goal is to achieve a very accurate emulation of a specific device and - thanks to the access to and understanding of the circuit schematic - deterministic methods are still favorable and the best performing, assuming flexibility of a model is not a requirement and development time is not a concern.

Non-differentiable gray-box approaches have also been widely studied, especially for nonlinear effects like amps, overdrive and distortion, thanks to the theoretical basis guaranteed by W-H models. These are also used in commercial products, although they have never been thoroughly evaluated in subjective tests; it is therefore difficult to know whether they can compete with white-box methods in terms of accuracy.

Non-differentiable black-box methods have also been studied, but the literature does not highlight the potential to achieve sufficient emulation accuracy while still being difficult to optimize.

The opposite can be said for differentiable methods, where thanks to the wide development of neural networks, black-box approaches have become the most studied and achieved state-of-the-art performance while differentiable white-box methods have not yet developed and need to be further explored to understand their performance potential.

Similarly to gray-box methods, the differentiable counterpart has been proved to be useful and worth exploring further even if, at the moment of writing, cannot achieve state-of-the-art emulation accuracy and have not been thoroughly evaluated in subjective tests.

\section{Methodology}
In the following sections we outline the key aspects of our work to establish a state of the art for differentiable black- and gray-box modeling. 
For consistency, ease of comparison, and to encourage future developments we rely on the \textbf{NablAFx}\footnote{\url{https://github.com/mcomunita/nablafx}}\citep{comunità2025nablafxframeworkdifferentiableblackbox} audio effects modeling framework.
We describe black- (Sec.~\ref{sec:bb-models}) and gray-box (Sec.~\ref{sec:gb-models}) architectures included in our experiments and introduce \textbf{ToneTwist AFx} (Sec.~\ref{sec:tonetwist}) - a novel dataset for audio effects research.

\subsection{Differentiable Black-box Models}
\label{sec:bb-models}
All architectures considered in our experiments are described in detail in \citep{comunità2025nablafxframeworkdifferentiableblackbox}; specifically, we include: LSTM-based recurrent networks, temporal convolution networks (TCN) \citep{lea2016temporal}, gated convolution networks (GCN) \cite{rethage2018wavenet} and structured state-space sequence models (S4) \citep{gu2021efficiently} with diagonal matrices \citep{gupta2022diagonal}.
Single LSTM layer architectures have been widely adopted for nonlinear (overdrive, distortion, guitar amps) \citep{wright2019real, wright2020real} and nonlinear time-varying (fuzz, compressor) \citep{steinmetz2022efficient, comunita2023modelling} effects.
Shown to outperform recurrent networks on various tasks, TCNs have also been adopted to model nonlinear time-varying effects, such as compressors \citep{steinmetz2022efficient}, while GCNs were proposed in \citep{
damskagg2019deep, 
damskagg2019real, 
wright2020real} 
for nonlinear (guitar amp, overdrive, distortion) and in \citep{comunita2023modelling} for nonlinear time-varying effects (compressor, fuzz).
More recently, S4 architectures were adopted for compressor modeling in \citep{yin2024modeling, simionato2024modeling}.
Since these architectures have each been applied to some but not all types of nonlinear effects, preventing clear conclusions on the state of the art, we include all of them in our experiments.

\subsection{Differentiable Gray-box Models}
\label{sec:gb-models}
As described in \citep{comunità2025nablafxframeworkdifferentiableblackbox}, we define a gray-box model as a sequence of differentiable processors, each with an associated controller which generates the control parameters that dictate the processor's exact behavior.
We propose gray-box architectures for dynamic range compression (Sec.\ref{sec:gb-comp}), distortion and fuzz (Sec.\ref{sec:gb-dist}).

\subsubsection{Model for Compressor/Limiter}
\label{sec:gb-comp}

While many approaches have been proposed to model dynamic range compression with differentiable black-box models 
\citep{
ramirez2019general, 
hawley2019profiling, 
steinmetz2022efficient, 
simionato2022deep,
comunita2023modelling, 
yeh2024hyper}
, there are fewer studies that adopt a gray-box paradigm 
\citep{
yu2024differentiable, 
colonel2022approximating, 
wright2022grey} 
and are all based on two previously suggested implementations of feed-forward digital compressors 
\cite{zolzer2002dafx, giannoulis2012digital}.

\begin{figure*}[t]
    \begin{minipage}[b]{\textwidth}
        \centering
        \includegraphics[width=.75\textwidth]{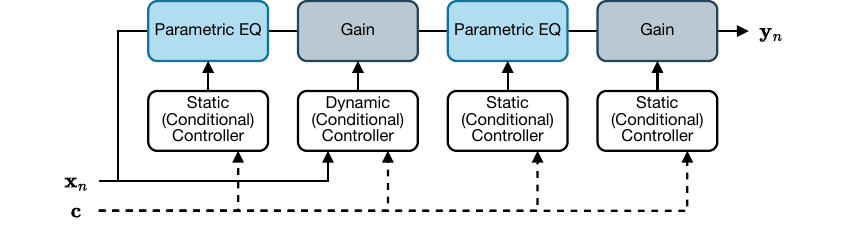}
        \\\textbf{(a)} Gray-box model for compressor/limiter
        \label{fig:gb-comp}
    \end{minipage}%
    \vspace{5mm}
    \begin{minipage}[b]{1\textwidth}
        \centering
        \includegraphics[width=\textwidth]{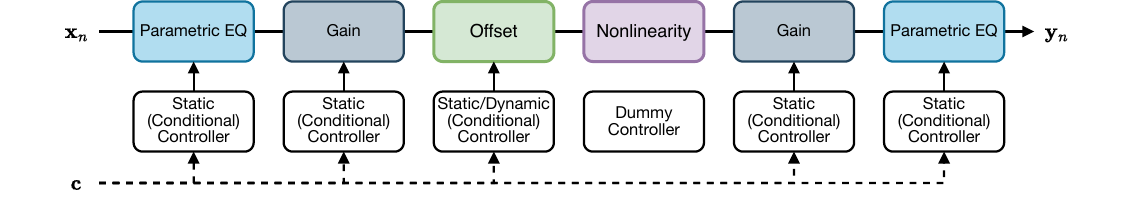}
        \\\textbf{(b)} Gray-box model for amp/overdrive/distortion/fuzz
        \label{fig:gb-dist}
    \end{minipage}
    \caption{Proposed gray-box architectures}
    \label{fig:gb-comp_gb-dist}
\end{figure*}

Here we propose a simple architecture leveraging dynamic controllers \citep{comunità2025nablafxframeworkdifferentiableblackbox} to implicitly learn the compression curve as a function of input signal and attack/release time constants.
The model - shown in Fig. \ref{fig:gb-comp_gb-dist} with input $x_{n}$, output $y_{n}$ and controls $c$ - is composed of: Static Parametric EQ, Dynamic Gain, Static Parametric EQ, Static Gain. 
The first Parametric EQ captures the overall frequency response of a device while the Dynamic Gain models the compression curve.
The second Parametric EQ shapes the output timbre after nonlinear gain, and the output Gain is necessary to match the target level (make-up gain).
The same model can be made parametric by replacing all static/dynamic controllers with their conditional counterparts.
The dynamic controller works at the default block size of 128 samples, equivalent to 2.66ms at 48kHz sampling rate. 
To obtain a higher temporal resolution, and more accurate modeling, one could use a smaller block size and implement interpolation methods for smoother control sequences.

In Fig. \ref{fig:gb-comp-dist_example-response} we show examples of the frequency/time response learned by the proposed model when trained on a compressor (Flamma AnalogComp) and a limiter (UA 6176 Vintage Channel Strip - 1176LN Limiter). 
In the first example, input and output EQs differ, while in the second, a nearly symmetrical response is learned.
We notice how the Dynamic Gain successfully captures different behaviors, with the compressor having slower attack and faster release than the limiter; also the exponential time constants - typical of analog compressors - appear to be correctly captured.

\subsubsection{Model for Amplifier, Overdrive, Distortion and Fuzz}
\label{sec:gb-dist}

Similar to compression, there is also a wealth of literature on differentiable black-box models for: 
amplifiers,
\citep{
covert2013vacuum, 
zhang2018vacuum, 
schmitz2018real, 
schmitz2018nonlinear, 
schmitz2019objective, 
damskagg2019deep, 
wright2019real, 
wright2020real, 
wright2023adversarial} 
overdrive 
\citep{
mendoza2005emulating, 
ramirez2019modeling, 
damskagg2019real, 
wright2020real, 
chowdhury2020comparison, 
fasciani2024conditioning, 
yeh2024hyper, 
simionato2024comparative}, 
distortion 
\citep{
ramirez2019modeling, 
damskagg2019real, 
wright2019real, 
wright2020real, 
yoshimoto2020deep, 
yoshimoto2021wavenet} 
and fuzz 
\citep{comunita2023modelling}. 
Recently, there has been exploration into applying a gray-box paradigm to this task 
\citep{
kuznetsov2020differentiable, 
nercessian2021lightweight, 
colonel2022reverse, 
miklanek2023neural, 
yeh2024ddsp}, 
as it offers desirable characteristics such as fewer trainable parameters, greater interpretability, and reduced training data requirements.

\begin{figure*}[t]
    \begin{minipage}[b]{.31\textwidth}
        \centering
        \includegraphics[height=3.7cm]{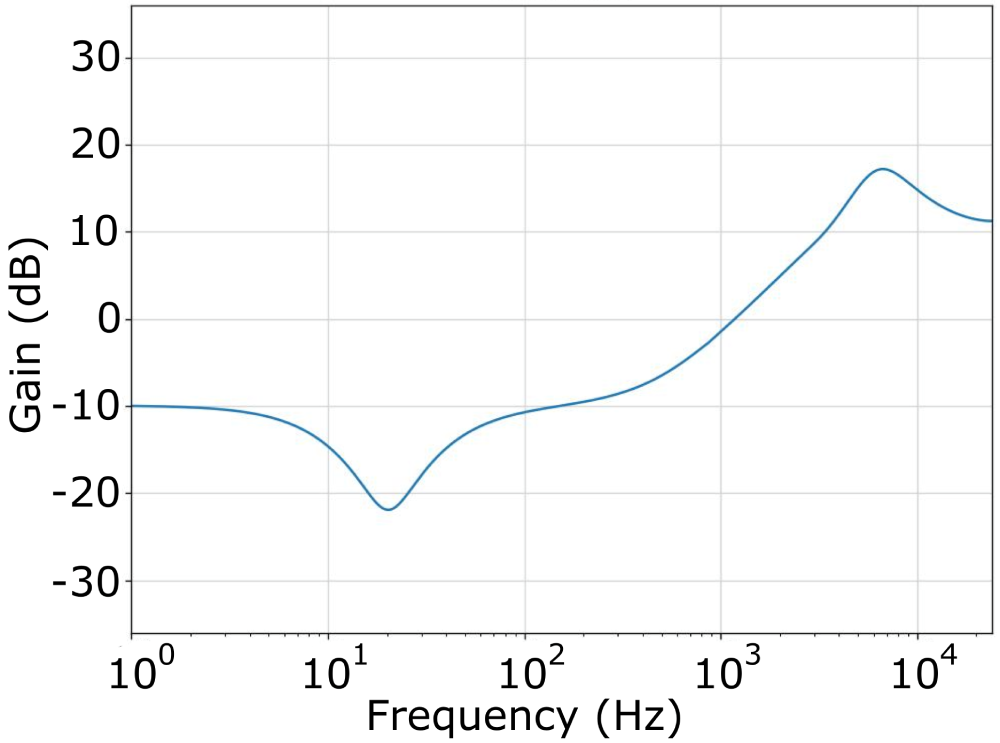}
        \\\textbf{(1a)} Parametric EQ
        \label{fig:ex_eqblock_response}
    \end{minipage}
    \begin{minipage}[b]{.36\textwidth}
        \centering
        \includegraphics[height=3.7cm]{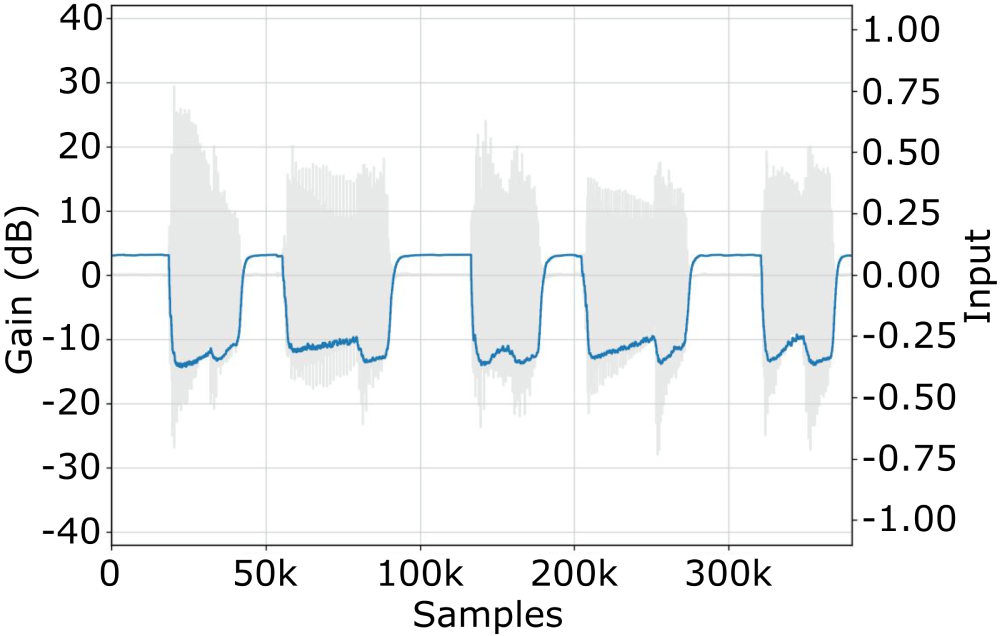}
        \\\textbf{(1b)} Dynamic Gain
        \label{fig:ex_nonlinblock_response}
    \end{minipage}
    \begin{minipage}[b]{.32\textwidth}
        \centering
        \includegraphics[height=3.7cm]{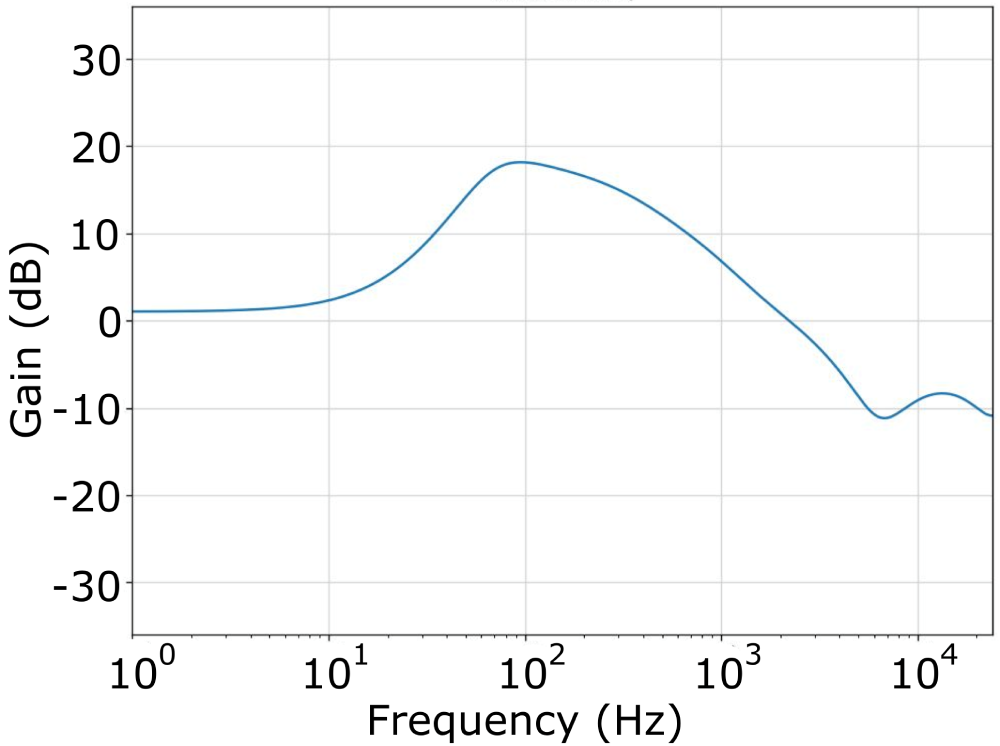}
        \\\textbf{(1c)} Parametric EQ
        \label{fig:ex_dcblock_response}
    \end{minipage}\\
    
    \begin{minipage}[b]{.31\textwidth}
        \centering
        \includegraphics[height=3.7cm]{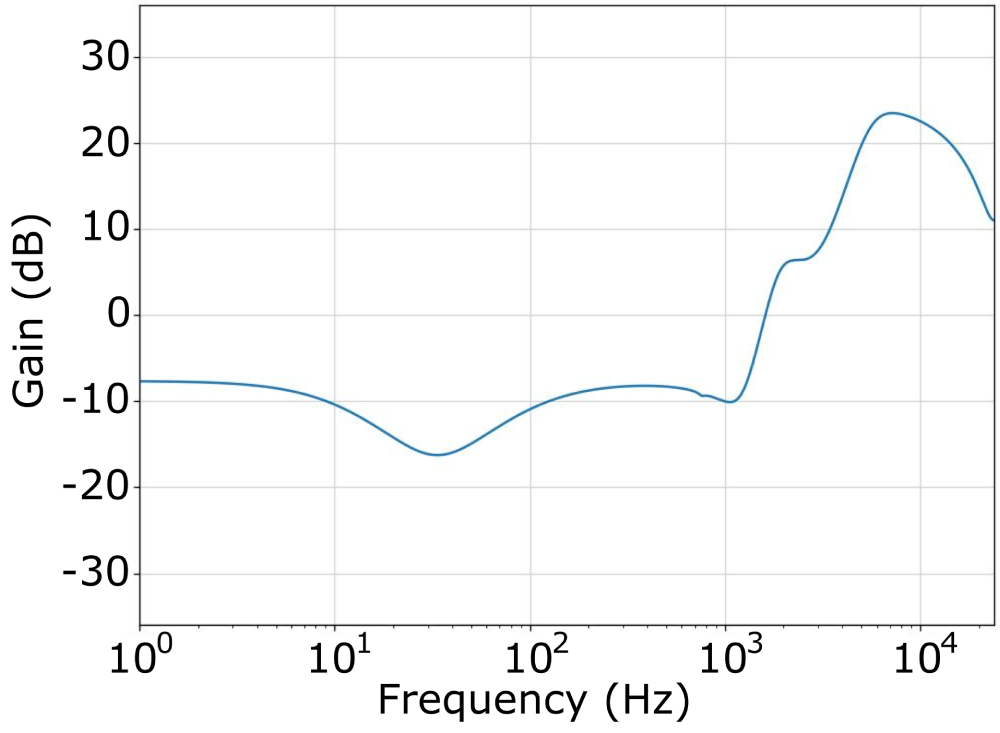}
        \\\textbf{(2a)} Parametric EQ
        \label{fig:ex_eqblock_response}
    \end{minipage}
    \begin{minipage}[b]{.36\textwidth}
        \centering
        \includegraphics[height=3.7cm]{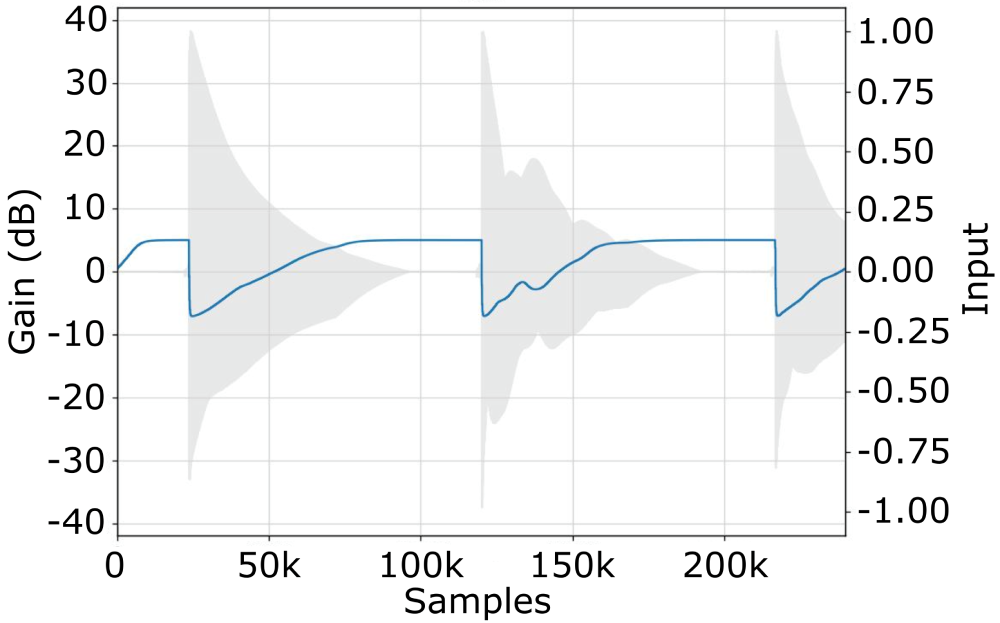}
        \\\textbf{(2b)} Dynamic Gain
        \label{fig:ex_nonlinblock_response}
    \end{minipage}
    \begin{minipage}[b]{.32\textwidth}
        \centering
        \includegraphics[height=3.7cm]{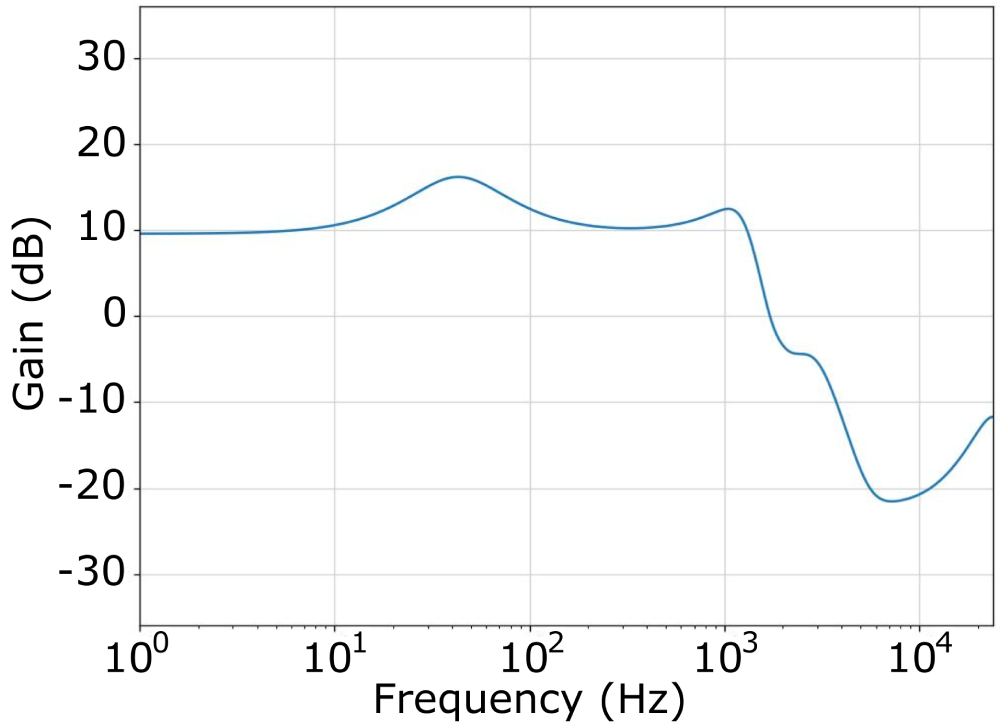}
        \\\textbf{(2c)} Parametric EQ
        \label{fig:ex_dcblock_response}
    \end{minipage}\\

    \begin{minipage}[b]{.50\textwidth}
        \centering
        \includegraphics[height=3.7cm]{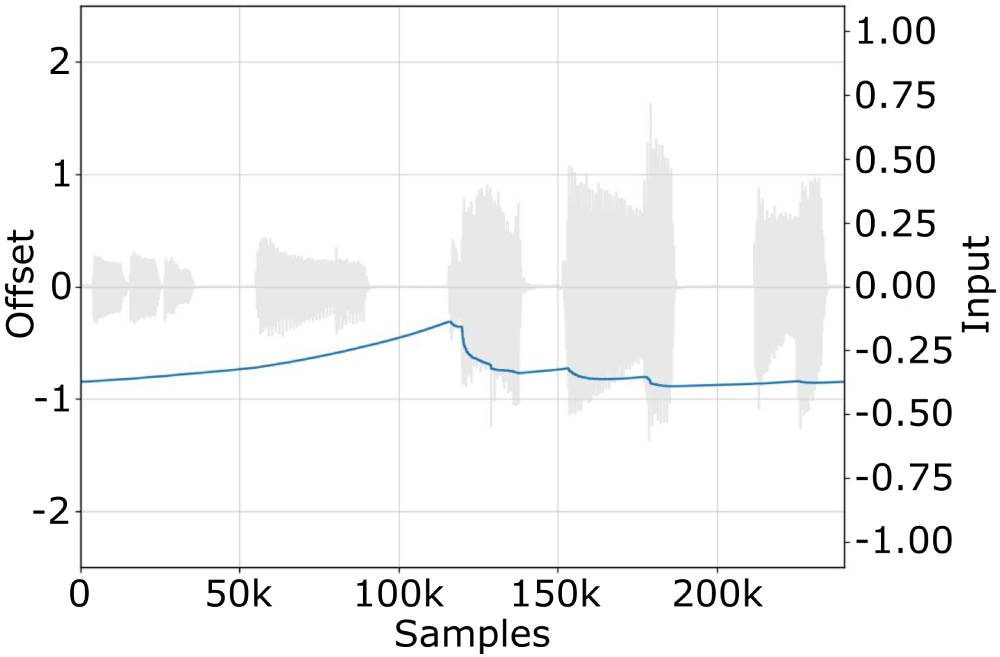}
        \\\textbf{(3a)} Dynamic Offset
        \label{fig:ex_offblock_response_1}
    \end{minipage}
    \begin{minipage}[b]{.50\textwidth}
        \centering
        \includegraphics[height=3.7cm]{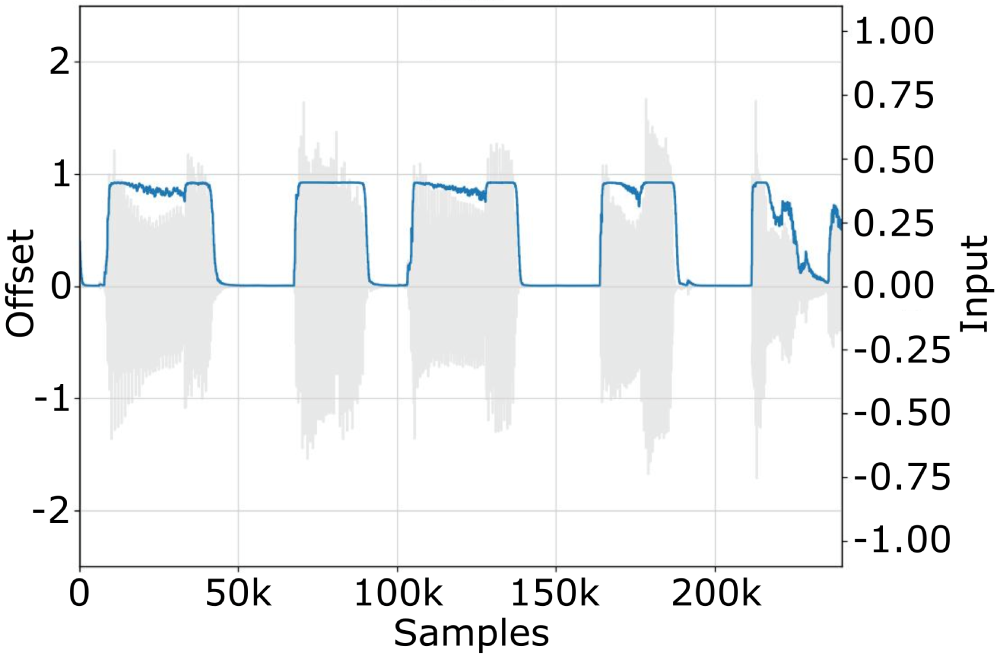}
        \\\textbf{(4a)} Dynamic Offset
        \label{fig:ex_offblock_response_2}
    \end{minipage}
    \caption{Examples of response for gray-box models of (1) Flamma Compressor, (2) UA 6176 - 1176LN Limiter, (3) Custom Dynamic Fuzz, (4) Harley Benton Fuzzy Logic.}
    \label{fig:gb-comp-dist_example-response}
\end{figure*}

In this work, we extend previous gray-box approaches based on the W-H model. 
Fig.~\ref{fig:gb-comp_gb-dist} illustrates our comprehensive architecture for emulating amplifiers, overdrive, distortion, and - when using dynamic controllers - fuzz effects.
We use two Parametric EQs \citep{comunità2025nablafxframeworkdifferentiableblackbox} for pre- and post-emphasis filtering around a Nonlinearity, which can be memoryless or not. 
Gain stages control distortion and output amplitude, while an Offset stage before the nonlinearity models asymmetrical clipping. 
A fixed offset suffices for amplifiers, overdrive, and distortion, while a dynamic controller is needed for fuzz effects to capture hysteresis (i.e., dynamic bias shift) and the characteristic timbre during note attack and release.
Conditional controllers can turn the model into a parametric one.
Fig.~\ref{fig:gb-comp-dist_example-response} shows examples of learned dynamic offsets from fuzz effects models, with the first having a longer attack and much longer release time constants than the second.

\subsection{ToneTwist AFx Dataset}
\label{sec:tonetwist}

In order to establish the state of the art in audio effects modeling across various effect types, we required access to dry-input/wet-output data from multiple devices; and, since modeling mainly focuses on physical units, data needed to be from analog devices.
For a realistic scenario, data needed to include a variety of sources, content and playing techniques, to test generalization capabilities beyond the ones used for training.
Furthermore, to run extensive experiments efficiently, we needed consistently organized data to avoid custom preprocessing and dataset code for each device.

Although there are publicly available datasets for audio effects modeling \citep{stein2010automatic, comunita2021guitar, pedroza2022egfxset}, these were deemed unsuitable for our work for several reasons.
The IDMT-SMT-Audio-Effects\footnote{\url{https://www.idmt.fraunhofer.de/en/publications/datasets/audio\_effects.html}} \citep{stein2010automatic} dataset focuses on detection and classification tasks, it is limited to single notes and chords from two electric guitars and two electric basses, and it only includes digital effects covering 10 linear, non-linear, and modulation effects.
Similarly, GUITAR-FX-DIST \citep{comunita2021guitar}, which focuses on classification and parameter estimation,  
includes single notes and chords from two guitars and only features digital emulations of 14 distortion effects.
While EGFxSet\footnote{\url{https://egfxset.github.io/}} \citep{pedroza2022egfxset} includes recordings from 12 physical units, equally split between analog and digital, it only features single notes from one electric guitar; and, despite being intended for audio effects modeling, it is unsuitable due to high background noise and other artifacts.

For the reasons above, we introduce the ToneTwist AFx dataset\footnote{\url{https://github.com/mcomunita/tonetwist-afx-dataset}} (Table~\ref{tab:dataset}), the largest collection (40 devices at the time of writing) of dry/wet signal pairs from nonlinear audio effects.
Beside being the first thoroughly documented audio effects dataset, it is also open to contributions from the wider community.
While data is hosted on Zenodo\footnote{\url{https://zenodo.org/}} for long-term availability, 
our repository organizes download links, provides retrieval scripts, basic training code, and contribution guidelines, facilitating new submissions and fostering discussion.

ToneTwist AFx is organized into four categories: analog, analog parametric, digital, and digital parametric, with \textit{parametric} referring to entries sampled with control settings granularity sufficient for parametric modeling.
It features dry-input signals from seven sources, including a variety of guitars, basses, and test signals like chirps and white noise.
The test set features different content and instruments from the training/validation set, ensuring a realistic evaluation of the generalization capabilities of trained models.

Recording was carried out with a Focusrite Scarlett Solo audio interface, featuring a low-noise pre-amplifier (±0.06 dB, 20Hz-20kHz). 
Synchronization between dry-input and wet-output was achieved by adding two impulses at the start and end of each file and then manually shifting each wet-output file of the same number of samples to account for interface latency.
Devices delays were assumed negligible.
We also ensure a high input dynamic range by applying a uniformly distributed gain ([-20dB, 0dB]) to the dry-input every 5 seconds. 

\renewcommand{\arraystretch}{1.15}
\begin{table*}[]
    \centering
    \caption{ToneTwist AFx Dataset (at the time of writing). For accessibility, we also include data from previous works (see Source) which we verify, pre-process (when necessary), rename, and reorganize for consistency.}
    \centerline{
        \begin{tabular}{lcccccc} 
            \midrule
            \midrule
            Type
                & Name 
                    & Analog/Digital
                        & Parametric    
                            & Source \\
            \midrule
            Amp 
                & Blackstar HT1 - Ch. Overdrive 
                & A 
                & N 
                & \cite{wright2019real} \\
            Amp 
                & Blackstar HT5 Metal - Ch. Overdrive 
                & A 
                & N 
                & \cite{schmitz2018introducing} \\
            Amp 
                & Engl Retro Tube 50 - Ch. Drive 
                & A 
                & N 
                & \cite{schmitz2018introducing} \\
            Amp 
                & Fender Blues Jr 
                & A 
                & N 
                & \citep{guitarml} \\
            Amp 
                & Ibanez TSA15 
                & A 
                & N 
                & \cite{schmitz2018introducing} \\
            Amp 
                & Marshall JVM410H - Ch. OD1 
                & A 
                & Y 
                & \cite{miklanek2023neural} \\
            Amp 
                & Mesa Boogie 5 50 Ch. Clean 
                & A 
                & N 
                & \cite{schmitz2018introducing} \\
            Amp 
                & Mesa Boogie 5 50 Ch. Crunch 
                & A 
                & N 
                & \cite{schmitz2018introducing} \\
            Amp 
                & Mesa Boogie 5 50 Ch. Burn 
                & A 
                & N 
                & \cite{schmitz2018introducing} \\
            Amp 
                & Mesa Boogie Mark V - Ch. Clean 
                & A 
                & N 
                & \cite{schmitz2018introducing} \\
            Amp 
                & Mesa Boogie Mark V - Ch. Crunch 
                & A 
                & N 
                & \cite{schmitz2018introducing} \\
            Amp 
                & Mesa Boogie Mark V - Ch. Extreme 
                & A 
                & N 
                & \cite{schmitz2018introducing} \\
            Pre-Amp 
                & UA 6176 Vintage Channel Strip - 610B 
                & A 
                & N 
                & \cite{martinez2020deep} \\
            \midrule
            Chorus 
                & Landlord Brewers Droop Chorus 
                & A 
                & N 
                & Proposed \\
            \midrule
            Compressor 
                & Ampeg Optocomp 
                & A 
                & N 
                & Proposed \\
            Compressor 
                & Amplitube DComp 
                & D 
                & N 
                & Proposed \\
            Compressor 
                & Amplitube Fender Comp 
                & D 
                & N 
                & Proposed \\
            Compressor 
                & Flamma AnalogComp 
                & A 
                & N 
                & Proposed \\
            Compressor 
                & Yuer DynaCompressor 
                & A 
                & N 
                & Proposed \\
            Limiter 
                & UA 6176 Vintage Channel Strip - 1176LN 
                & A 
                & N 
                & \cite{martinez2020deep} \\
            \midrule
            Overdrive 
                & DIY Klon Centaur 
                & A 
                & N 
                & \cite{guitarml} \\
            Overdrive 
                & Fulltone Fulldrive 2 
                & A 
                & N 
                & Proposed \\
            Overdrive 
                & Ibanez TS9 
                & A 
                & N 
                & \cite{guitarml} \\
            Overdrive 
                & Harley Benton Green Tint 
                & A 
                & N 
                & Proposed \\
            Overdrive 
                & Multidrive Pedal Pro 808-Scream 
                & D 
                & Y 
                & Proposed \\
            Overdrive 
                & Multidrive Pedal Pro B-Drive 
                & D 
                & Y 
                & Proposed \\
            \midrule
            Distortion 
                & DIY Electro Harmonix Big Muff 
                & A 
                & N 
                & \cite{guitarml} \\
            Distortion 
                & Electro Harmonix Big Muff 
                & A 
                & N 
                & \cite{wright2019real} \\
            Distortion 
                & Electro Harmonix Metal Muff 
                & A 
                & N 
                & Proposed \\
            Distortion 
                & Harley Benton Big Fur 
                & A 
                & N 
                & Proposed \\
            Distortion 
                & Harley Benton DropKick 
                & A 
                & N 
                & Proposed \\
            Distortion 
                & Harley Benton Plexicon 
                & A 
                & N 
                & Proposed \\
            Distortion 
                & Harley Benton Rodent 
                & A 
                & N 
                & Proposed \\
            Distortion 
                & Multidrive Pedal Pro M-Distortion 
                & D 
                & Y 
                & Proposed \\
            \midrule
            Fuzz 
                & Custom Dynamic Fuzz 
                & D 
                & N 
                & Proposed \\
            Fuzz 
                & Harley Benton Fuzzy Logic 
                & A 
                & N 
                & Proposed \\
            Fuzz 
                & Harley Benton Silly Fuzz 
                & A 
                & N 
                & Proposed \\
            Fuzz 
                & Multidrive Pedal Pro F-Fuzz 
                & D 
                & Y 
                & Proposed \\
            Fuzz 
                & Arturia Spring636 Germanium Preamp 
                & D 
                & N 
                & Proposed \\
            \midrule
            Tremolo 
                & Mooer Trelicopter 
                & A 
                & N 
                & Proposed \\
            \midrule
            \midrule
        \end{tabular}
    }
    \label{tab:dataset}
\end{table*}





\section{Experiments}
\label{sec:experiments}
In this section we give an overview and description of all the architectures, models configurations, and data used for our experiments.
To help establish the state of the art in audio effects modeling and identify future research directions, we included a wide range of experiments and configurations.

\renewcommand{\arraystretch}{0.95}
\begin{table*}[h!]
    \caption{Configurations for models included in the experiments. \\
    PEQ=Parametric EQ, G=Gain, O=Offset, MLP=Multilayer Perceptron, RNL=Rational Non Linearity.\\
    .s=static, .d=dynamic, .sc=static conditional, .dc=dynamic conditional}
    \label{tab:models} 
    \centerline{
        \begin{tabular}{L{3.0cm}C{1.1cm}R{1.5cm}C{1.0cm}C{1.0cm}C{1.2cm}C{1.4cm}R{1.3cm}R{1.2cm}R{1.2cm}}
            \midrule
            \midrule
            Model
                & Cond.
                    & R.F. (samples)
                        & Blocks
                            & Kernel
                                & Dilation
                                    & Channels
                                        & \# Params 
                                            & FLOP/s 
                                                & MAC/s\\ 
            \midrule
            LSTM-32 & - & - & 1 & - & - & 32 & 4.5k & 423.9M & 1.5M\\
            LSTM-96 & - & - & 1 & - & - & 96 & 38.1k & 3.63G & 4.6M\\
            \midrule
            TCN-45-S-16 & - & 2047 & 5 & 7 & 4 & 16 & 7.5k & 732.5M & 364.2M\\
            TCN-45-L-16 & - & 2047 & 10 & 3 & 2 & 16 & 7.3k & 715.9M & 353.9M\\
            TCN-250-S-16 & - & 13651 & 6 & 11 & 4 & 16 & 14.5k & 1.67G & 834.3M\\
            TCN-250-L-16 & - & 12283 & 11 & 7 & 2 & 16 & 18.4k & 2.12G & 1.05G\\
            TCN-2500-S-16 & - & 133333 & 5 & 13 & 10 & 16 & 13.7k & 3.96G & 1.97G\\
            TCN-2500-L-16 & - & 118097 & 10 & 5 & 3 & 16 & 11.9k & 3.48G & 1.73G\\
            \midrule
            TCN-TF-45-S-16 & TFiLM & 2047 & 5 & 7 & 4 & 16 & 39.5k & 760.8M & 364.2M\\
            TCN-TF-45-L-16 & TFiLM & 2047 & 10 & 3 & 2 & 16 & 71.3k & 772.7M & 353.9M\\
            TCN-TF-250-S-16 & TFiLM & 13651 & 6 & 11 & 4 & 16 & 52.9k & 1.72G & 834.3M\\
            TCN-TF-250-L-16 & TFiLM & 12283 & 11 & 7 & 2 & 16 & 88.8k & 2.19G & 1.05G\\
            TCN-TF-2500-S-16 & TFiLM & 133333 & 5 & 13 & 10 & 16 & 45.7k & 4.05G & 1.97G\\
            TCN-TF-2500-L-16 & TFiLM & 118097 & 10 & 5 & 3 & 16 & 75.9k & 3.65G & 1.73G\\
            \midrule
            GCN-45-S-16 & - & 2047 & 5 & 7 & 4 & 16 & 16.2k & 1.58G & 786.2M\\
            GCN-45-L-16 & - & 2047 & 10 & 3 & 2 & 16 & 17.1k & 1.67G & 825.1M\\
            GCN-250-S-16 & - & 13651 & 6 & 11 & 4 & 16 & 30.4k & 3.52G & 1.75G\\
            GCN-250-L-16 & - & 12283 & 11 & 7 & 2 & 16 & 39.6k & 4.55G & 2.26G\\
            GCN-2500-S-16 & - & 133333 & 5 & 13 & 10 & 16 & 28.6k & 8.27G & 4.12G\\
            GCN-2500-L-16 & - & 118097 & 10 & 5 & 3 & 16 & 26.4k & 7.65G & 3.79G\\
            \midrule
            GCN-TF-45-S-16 & TFiLM & 2047 & 5 & 7 & 4 & 16 & 141.6k & 1.69G & 786.2M\\
            GCN-TF-45-L-16 & TFiLM & 2047 & 10 & 3 & 2 & 16 & 268.0k & 1.88G & 825.1M\\
            GCN-TF-250-S-16 & TFiLM & 13651 & 6 & 11 & 4 & 16 & 181.0k & 3.67G & 1.75G\\
            GCN-TF-250-L-16 & TFiLM & 12283 & 11 & 7 & 2 & 16 & 315.6k & 4.82G & 2.26G\\
            GCN-TF-2500-S-16 & TFiLM & 133333 & 5 & 13 & 10 & 16 & 154.1k & 8.59G & 4.12G\\
            GCN-TF-2500-L-16 & TFiLM & 118097 & 10 & 5 & 3 & 16 & 277.3k & 8.27G & 3.79G\\
            \midrule
            \midrule
        \end{tabular}
    }
    \centerline{
        \begin{tabular}{L{3.0cm}C{1.1cm}R{1.5cm}C{1.0cm}C{2.6cm}C{1.4cm}R{1.3cm}R{1.2cm}R{1.2cm}}
            Model
                & Cond.
                    & R.F. (samples)
                        & Blocks
                            & State Dimension
                                & Channels
                                    & \# Params
                                        & FLOP/s 
                                            & MAC/s\\ 
            \midrule
            S4-S-16 & - & - & 4 & 4 & 16 & 2.4k & 132.1M & 53.8M\\
            S4-L-16 & - & - & 8 & 32 & 16 & 19.0k & 605.2M & 106.0M\\
            \midrule
            S4-TF-S-16 & TFiLM & - & 4 & 4 & 16 & 28.0k & 154.1M & 53.8M\\
            S4-TF-L-16 & TFiLM & - & 8 & 32 & 16 & 70.2k & 649.1M & 106.0M\\
            \midrule
            \midrule
        \end{tabular}
    }
    \centerline{
        \begin{tabular}{L{3cm}C{9.3cm}R{1.3cm}R{1.2cm}R{1.2cm}}
            Model
                & Signal Chain
                    & \# Params
                        & FLOP/s 
                            & MAC/s\\
            \midrule
            GB-COMP & PEQ.s $\rightarrow$ G.d $\rightarrow$ PEQ.s $\rightarrow$ G.s & 47 & 58k & 0\\
            \midrule
            GB-DIST-MLP & PEQ.s $\rightarrow$ G.s $\rightarrow$ O.s $\rightarrow$ MLP $\rightarrow$ G.s $\rightarrow$ PEQ.s & 2.2k & 202.8M & 101.4M\\
            GB-DIST-RNL & PEQ.s $\rightarrow$ G.s $\rightarrow$ O.s $\rightarrow$ RNL $\rightarrow$ G.s $\rightarrow$ PEQ.s & 47 & 912k & 0\\
            \midrule
            GB-FUZZ-MLP & PEQ.s $\rightarrow$ G.s $\rightarrow$ O.d $\rightarrow$ MLP $\rightarrow$ G.s $\rightarrow$ PEQ.s & 2.3k & 202.8M & 101.4M\\
            GB-FUZZ-RNL & PEQ.s $\rightarrow$ G.s $\rightarrow$ O.d $\rightarrow$ RNL $\rightarrow$ G.s $\rightarrow$ PEQ.s & 62 & 970k & 0\\
            \hline
            \hline
        \end{tabular}
    }
\end{table*}
\renewcommand{\arraystretch}{0.95}
\begin{table*}[h!]
    \caption{Configurations for parametric models included in the experiments. \\
    PEQ=Parametric EQ, G=Gain, O=Offset, MLP=Multilayer Perceptron, RNL=Rational Non Linearity.\\
    .s=static, .d=dynamic, .sc=static conditional, .dc=dynamic conditional}
    \label{tab:models-param}
    \centerline{
        \begin{tabular}{L{3.0cm}C{1.5cm}R{1.5cm}C{1.0cm}C{1.0cm}C{1.2cm}C{1.4cm}R{1.3cm}R{1.2cm}R{1.2cm}}
            \midrule
            \midrule
            Model
                & Cond.
                    & R.F. (samples)
                        & Blocks
                            & Kernel
                                & Dilation
                                    & Channels
                                        & \# Params 
                                            & FLOP/s 
                                                & MAC/s\\ 
            \midrule
            LSTM-C-32 & Concat & - & 1 & - & - & 32 & 5.0k & 473.1M & 1.5M\\
            LSTM-TVC-32 & TVConcat & - & 1 & - & - & 32 & 8.0k & 621.7M & 1.5M\\
            \midrule
            LSTM-C-96 & Concat & - & 1 & - & - & 96 & 39.7k & 3.78G & 4.6M\\
            LSTM-TVC-96 & TVConcat & - & 1 & - & - & 96 & 45.7k & 4.22G & 4.6M\\
            \midrule
            \midrule
            TCN-F-45-S-16 & FiLM & 2047 & 5 & 7 & 4 & 16 & 15.0k & 736.5M & 364.3M\\
            TCN-TF-45-S-16 & TFiLM & 2047 & 5 & 7 & 4 & 16 & 42.0k & 762.8M & 364.2M\\
            TCN-TTF-45-S-16 & TTFiLM & 2047 & 5 & 7 & 4 & 16 & 17.3k & 744.0M & 367.4M\\
            TCN-TVF-45-S-16 & TVFiLM & 2047 & 5 & 7 & 4 & 16 & 17.7k & 740.4M & 366.2M\\
            \midrule
            TCN-F-45-L-16 & FiLM & 2047 & 5 & 7 & 4 & 16 & 20.1k & 723.8M & 354.0M\\
            TCN-TF-45-L-16 & TFiLM & 2047 & 5 & 7 & 4 & 16 & 76.4k & 776.7M & 353.9M\\
            TCN-TTF-45-L-16 & TTFiLM & 2047 & 5 & 7 & 4 & 16 & 27.0k & 739.0M & 360.4M\\
            TCN-TVF-45-L-16 & TVFiLM & 2047 & 5 & 7 & 4 & 16 & 22.8k & 727.9M & 357.9M\\
            \midrule
            \midrule
        \end{tabular}
    }
    \centerline{
        \begin{tabular}{L{3.0cm}C{1.5cm}R{1.5cm}C{1.0cm}C{2.6cm}C{1.4cm}R{1.3cm}R{1.2cm}R{1.2cm}}
            Model
                & Cond.
                    & R.F. (samples)
                        & Blocks
                            & State Dimension
                                & Channels
                                    & \# Params
                                        & FLOP/s 
                                            & MAC/s\\ 
            \midrule
            S4-F-S-16 & FiLM & - & 4 & 4 & 16 & 8.9k & 135.2M & 53.8M\\
            S4-TF-S-16 & TFiLM & - & 4 & 4 & 16 & 30.0k & 155.6M & 53.8M\\
            S4-TTF-S-16 & TTFiLM & - & 4 & 4 & 16 & 10.2k & 141.0M & 56.3M\\
            S4-TVF-S-16 & TVFiLM & - & 4 & 4 & 16 & 11.6k & 138.9M & 55.3M\\
            \midrule
            S4-F-L-16 & FiLM & - & 8 & 32 & 16 & 29.7k & 611.3M & 106.0M\\
            S4-TF-L-16 & TFiLM & - & 8 & 32 & 16 & 74.3k & 652.2M & 105.9M\\
            S4-TTF-L-16 & TTFiLM & - & 8 & 32 & 16 & 34.8k & 623.1M & 111.0M\\
            S4-TVF-L-16 & TFiLM & - & 8 & 32 & 16 & 32.4k & 615.1M & 109.1M\\
            \midrule
            \midrule
        \end{tabular}
    }
    \centerline{
        \begin{tabular}{L{3cm}C{9.7cm}R{1.3cm}R{1.2cm}R{1.2cm}}
            Model
                & Signal Chain
                    & \# Params
                        & FLOP/s 
                            & MAC/s\\
            \midrule
            GB-C-DIST-MLP & PEQ.sc $\rightarrow$ G.sc $\rightarrow$ O.sc $\rightarrow$ MLP $\rightarrow$ G.sc $\rightarrow$ PEQ.sc & 4.5k & 202.8M & 101.4M\\
            GB-C-DIST-RNL & PEQ.sc $\rightarrow$ G.sc $\rightarrow$ O.sc $\rightarrow$ RNL $\rightarrow$ G.sc $\rightarrow$ PEQ.sc & 2.3k & 920.5k & 4.3k\\
            \midrule
            GB-C-FUZZ-MLP & PEQ.sc $\rightarrow$ G.sc $\rightarrow$ O.dc $\rightarrow$ MLP $\rightarrow$ G.sc $\rightarrow$ PEQ.sc & 4.2k & 202.8M & 101.4M\\
            GB-C-FUZZ-RNL & PEQ.sc $\rightarrow$ G.sc $\rightarrow$ O.dc $\rightarrow$ RNL $\rightarrow$ G.sc $\rightarrow$ PEQ.sc & 2.0k & 988.9k & 3.6k\\
            \midrule
            \midrule
        \end{tabular}
    }
\end{table*}

\subsection{Overview of Models Configurations}
\label{sec:models_config}
Table \ref{tab:models} lists all models, hyperparameter values, and computational complexity\footnote{\url{https://github.com/MrYxJ/calculate-flops.pytorch}} used in our experiments on non-parametric modeling, where we balance tractability by limiting the number of models per architecture while covering a broad range of sizes and complexities.

As described in Sec.\ref{sec:bb-models}, we include the four most common neural networks types for black-box modeling: LSTM, TCN, GCN and S4; as well as the gray-box models proposed in Sec.\ref{sec:gb-models}: GB-COMP for compressor/limiter, GB-DIST for amplifier/overdrive/distortion and GB-FUZZ for fuzz.
For LSTM architectures, the only parameter to choose is the number of channels, with 32 and 96 channels being sensible choices for \textit{small} and \textit{large} models (LSTM-32/LSTM-96), as adopted in previous works \cite{wright2020real, steinmetz2022efficient, comunita2023modelling}.\\
For TCN and GCN models, we consider variants with \textit{short}, \textit{medium}, and \textit{long} receptive fields (45, 250 and 2500~ms at 48kHz), with \textit{small} (5/6 blocks) or \textit{large} (10/11 blocks) configurations (S/L in model names).
We keep the number of channels constant and equal to 16.
We also include variants without and with TFiLM (TF in the models' names) to evaluate whether time-varying conditioning - shown to be able to capture long-range dependencies \citep{comunita2023modelling} typical of compressors and fuzz effects - is beneficial across architectures. 
For TFiLM layers we always adopt a block size - i.e., downsampling/maxpooling factor - of 128, which was shown to be optimal in \citep{comunita2023modelling}.
As shown in the table, for each receptive field, \textit{small} and \textit{large} models have comparable parameters, FLOP/s, and MAC/s, with the idea that the main difference is in the number of processing stages and not in the networks expressivity itself.
With TFiLM adding a conditioning layer per block, matching parameter counts across block variations isn't feasible, but due to the small channel count (16) and low TFiLM sample rate, model pairs remain computationally comparable.\\
For S4 architectures we define \textit{small} and \textit{large} models (S/L in the name) based on the number of blocks and state dimension, with values chosen based on the experiments reported in the original publication \citep{gu2021efficiently}.
For S4 models as well we introduce TFiLM conditioned variants and keep the number of channels constant and equal to 16.\\
While for the GB-COMP architecture we only investigate one configuration, for GB-DIST and GB-FUZZ models we test two, based on the type of nonlinearity adopted: Static MLP Nonlinearity or Static Rational Nonlinearity (MLP/RNL in the name).
The MLP itself accounts for most of the computational complexity of GB-DIST and GB-FUZZ models.\\

For experiments on parametric modeling (Table~\ref{tab:models-param}) we take a subset of the configurations in Table~\ref{tab:models} and we pair them with the various conditioning mechanisms described in \citep{comunità2025nablafxframeworkdifferentiableblackbox}.
For LSTM models, we use concatenation (C) as the baseline and also test time-varying concatenation (TVC), which adds minimal computational cost.
For convolutional backbones, we select only TCNs and models with a \textit{short} (45ms) receptive field, as these rely more on conditioning methods to capture long-range dependencies, highlighting performance differences.
We consider FiLM conditioning (F in the name) to be the baseline method and experiment with TFiLM, TTFiLM (TTF in the name) and TVFiLM (TVF in the name). 
We do the same for S4 architectures.
Table~\ref{tab:models-param} shows that both TTFiLM and TVFiLM introduce time-dependent conditioning with parameter and computational efficiency similar to the baseline FiLM method, having minimal impact on parameters, FLOP/s, and MAC/s.
For gray-box models, we add parametric control using differentiable controllers from \citep{comunità2025nablafxframeworkdifferentiableblackbox} and replace static/dynamic controllers with conditional ones (C in the model name).
Although they increase the number of parameters, the impact of conditional controllers on the computational complexity is minimal due to the small size of the MLPs and LSTMs in their implementation.

Beside a good understanding of the state-of-the-art, our experiments aim to identify architectures that can model a wide range of effects and devices consistently, without requiring different configurations, with the broader goal of defining a universal audio effects modeling architecture.
While we recognize that model pruning \citep{sudholt2022pruning} or hyperparameter optimization could identify the smallest model size for each effect, this would require extensive time and resources. 
Alternatively, model distillation \citep{hinton2015distillingknowledgeneuralnetwork} could reduce size and increase efficiency after identifying a suitable architecture, but this also requires additional experimentation and training.

\subsection{Overview of Audio Effects Configurations}
Table~\ref{tab:effects_configs} lists all the audio effects and settings in our study. For non-parametric modeling, we select four devices per effect category (compressor/limiter, overdrive, distortion, fuzz), totaling sixteen devices, twelve of which are from our ToneTwist AFX dataset (see Sec.~\ref{sec:tonetwist}).

Settings for each device are chosen to ensure a variety of behaviors within each category, while maximizing the challenge for the modeling task.
For compression, we select fast attack, slow release, and high compression ratio settings to capture fast transients, long-range dependencies, and nonlinear behaviors. 
For overdrive, we choose high gain settings with varying tone/equalization for a complex spectrum.
For distortion, we select both \textit{medium} and \textit{high} gain settings, while for fuzz effects, which are inherently \textit{high} gain, we opt for \textit{medium} gain settings to ensure the dynamic behavior is noticeable and not concealed by excessive distortion.
It is important to note that the settings values across effect types and devices are only vaguely correlated with the behaviors we are modeling, as they don't correspond to specific units or internal parameters, serving only as a general description of the effect's behavior and timbral traits.

The table also includes effects and settings for parametric modeling experiments: a vacuum tube guitar amplifier and a digital germanium transistor fuzz emulation.
In addition to parametric modeling and conditioning methods, data from these two devices help evaluate performance on unseen control configurations and sources. 
For the Marshall JVM410H, training, validation, and testing use the same sources but different control settings. 
For the Multidrive Pedal Pro F-Fuzz, control settings are the same across sets, but sources and content differ, featuring unseen guitar and bass performances.

\begin{table*}[t]
    \centering
    \caption{Audio effects configurations use in the experiments. For parametric models not all combinations are part of the data.}
    \label{tab:effects_configs}
    \centerline{
        \begin{tabular}{L{2cm}L{6.5cm}l}
        \midrule
        \midrule
            Type & Name & Configuration for Train/Val/Test\\
            \midrule
            Compressor & Ampeg Optocomp & Compression: 10, Release: 10, Level: 6\\
            Compressor & Flamma AnalogComp & Volume: 10, Comp: 10, Eq: 5\\
            Compressor & Yler DynaCompressor & Attack: 5, Sustain: 10, Level: 10\\
            Limiter & UA 6176 Vintage Channel Strip - 1176LN & Attack: 10, Release: 10, Input: 4, Output: 7, Ratio: All\\
            \midrule
            Overdrive & DIY Klon Centaur & Diodes: Si, Gain: 9, Tone: 8, Out: 6\\
            Overdrive & Fulltone Fulldrive 2 & Volume: 10, Tone: 5, Overdrive: 10, Boost: Off\\
            Overdrive & Ibanez TS9 & Volume: 5, Tone: 5, Drive: 10\\
            Overdrive & Harley Benton Green Tint & Volume: 10, Tone: 10, Gain: 10\\
            \midrule
            Distortion & Electro Harmonix Big Muff & Sustain: 5, Volume: 10\\
            Distortion & Harley Benton DropKick & Volume: 10, Tone: 5, Gain: 10, Mode: Normal\\
            Distortion & Harley Benton Plexicon & Volume: 10, Tone: 5, Gain: 5, Mode: Normal\\
            Distortion & Harley Benton Rodent & Volume: 10, Filter: 5, Distortion: 10, Mode: Normal\\
            \midrule
            Fuzz & Custom Dynamic Fuzz & Gain: 5, Sensitivity: 10, Attack: 0, Release: 10, Volume: 10\\
            Fuzz & Harley Benton Fuzzy Logic & Volume: 10, Fuzz: 5\\
            Fuzz & Harley Benton Silly Fuzz & Volume: 10, Fuzz: 5\\
            Fuzz & Arturia Spring636 Germanium Preamp & Mode: Aux, Input Gain: 5, Blend: 0, Output Gain: 6\\
            \midrule
            \midrule
        \end{tabular}
    }
    \centerline{
        \begin{tabular}{L{2cm}L{4.4cm}ll}
            Type & Name & Configuration for Train/Val & Configuration for Test\\
            \midrule
            Amp & Marshall JVM410H - Ch. OD1 
                & Bass: [0, 2, 4, 5, 6, 8, 10] & Bass: [0, 1, 3, 6.5, 10]\\
                && Middle: [0, 2, 4, 5, 6, 8, 10] & Middle: [0, 1, 3, 7, 8.5, 10]\\
                && Treble: [0, 2, 4, 5, 6, 8, 10] & Treble: [0, 1, 3.5, 7, 10]\\
                && Gain: [0, 1, 2, 4, 5, 6, 8, 10] & Gain: [5]\\
            Fuzz & Multidrive Pedal Pro F-Fuzz
                & Fuzz: [0, 1, 2, 3, 4, 5, 6, 7, 8, 9, 10] & Fuzz: [0, 1, 2, 3, 4, 5, 6, 7, 8, 9, 10]\\
                && Volume: [10] & Volume: [10]\\
            \midrule
            \midrule
        \end{tabular}
    }
\end{table*}

\subsection{Training}
\label{sec:training}

\renewcommand{\arraystretch}{1.1}
\begin{table*}[b]
    \setlength{\tabcolsep}{0pt}
    \caption{Training configurations for each architecture. LR = learning rate. MR-STFT = multi-resolution spectrogram loss. TBPTT = truncated backpropagation through time.}
    \label{tab:training}
    \centerline{
        \begin{tabular}{L{4cm} C{2.5cm} C{2cm} C{2cm} C{2cm} C{2cm}}
            \midrule
            \midrule
            & LSTM & TCN & GCN & S4 & GB\\
            \hline
        \end{tabular}
    }
    \centerline{
        \begin{tabular}{L{4cm} C{2.5cm} C{2cm} C{2cm} C{2cm} C{2cm}}
            LR & [0.005, 0.001] & [0.005] & [0.005] & [0.01] & [0.1]\\
            \hline
        \end{tabular}
    }
    \centerline{
        \begin{tabular}{L{4cm} C{10.555cm}}
            LR Scheduler & Reduce on plateau (patience = 20 epochs, factor = 0.5)\\
            \hline
        \end{tabular}
    }
    \centerline{
        \begin{tabular}{L{4cm} C{10.555cm}}
            Early Stopping & Total validation loss (patience = 50 epochs)\\
            \hline
        \end{tabular}
    }
    \centerline{
        \begin{tabular}{L{4cm} C{10.555cm}}
            Loss weights (L1/MR-STFT) & [10/1, 5/5, 1/0.1, 0.5/0.5]\\
            \hline
        \end{tabular}
    }
    \centerline{
        \begin{tabular}{L{4cm} | C{2.5cm} | C{8cm}}
            Gradient clip alg. & Norm & Value\\
            \hline
        \end{tabular}
    }
    \centerline{
        \begin{tabular}{L{4cm} | C{2.5cm} | C{8cm}}
            Gradient clip value & 10 & 1\\
            \hline
        \end{tabular}
    }
    \centerline{
        \begin{tabular}{L{4cm} | C{10.52cm}}
            Sample length & 3~s\\
            \hline
        \end{tabular}
    }
    \centerline{
        \begin{tabular}{L{4cm} | C{10.52cm}}
            Train/Val split & 0.9/0.1\\
            \hline
        \end{tabular}
    }
    \centerline{
        \begin{tabular}{L{4cm} | C{2.5cm} | C{8cm}}
            Max steps & 451k & 15k\\
            \hline
        \end{tabular}
    }
    \centerline{
        \begin{tabular}{L{4cm} | C{2.5cm} | C{8cm}}
            TBPTT & Y & N\\
            \hline
        \end{tabular}
    }
    \centerline{
        \begin{tabular}{L{4cm} | C{2.5cm} | C{8cm}}
            Update weights every & 100~ms & 3~s\\
        \hline
        \hline
        \end{tabular}
    }
\end{table*}

All previous audio effects modeling studies have always assumed that training different architectures and model configurations under the same training scheme - i.e., loss functions and loss weights, learning rate and rate scheduler etc. - allows fair performance comparisons and reliable state-of-the-art conclusions.
This has been the case also for works that compared vastly different architectures like black-box and gray-box ones 
\citep{yeh2024ddsp, miklanek2023neural, wright2022grey}.
Although, in preliminary experiments we were able to assess that training an architecture under different schemes can yield vastly different loss values, and even more so when different architectures are involved.
This highlights the need to choose appropriate training-related hyperparameters for each model to enable meaningful comparisons between architectures and configurations.

This section discusses preliminary experiments to identify optimal hyperparameters and training configurations for LSTM, TCN, GNC, S4, and GB architectures, and the resulting training schemes are shown in Table~\ref{tab:training}.
Based on prior work on effects modeling \cite{steinmetz2022efficient, comunita2023modelling}, sound synthesis \cite{engel2019ddsp}, and audio processing \citep{yamamoto2020parallel}, we assumed an $L1$ + MR-STFT loss as the most reliable and perceptually relevant choice, and opted for specific combinations of loss weights, which we report in Table~\ref{tab:training}.
Having a 10:1 ratio between $L1$ and MR-STFT loss makes the two terms about equal in absolute numeric values, giving them the same importance.
Conversely, having a 1:1 ratio equals to give the spectrogram loss greater importance.
To compare experiments with different loss weights, we report results by scaling the loss terms accordingly.
This means that, while backpropagation is influenced by the weight values, the results in Section~\ref{sec:results} are independent of the specific choice, allowing for comparison across experiments.

We chose the learning rate (LR) for each architecture by training on one audio effect (Custom Dynamic Fuzz) and comparing results for $\text{LR}=[0.1,0.01,0.005,0.001,0.0005]$.
For gray-box models we also tested $LR=1$, since higher learning rates seemed to work better. 
We also found it essential to use a learning rate multiplier for each block in the gray-box signal chain, as more complex blocks (e.g., Static MLP Nonlinearity, Static FIR Filter) needed a lower LR than simpler ones (e.g., Parametric EQ, Static Rational Nonlinearity).
This significantly improved gray-box losses and led to better local optima in terms of perceptual similarity with the target.
Furthermore, we conducted experiments to choose activation functions and, after testing $tanh$, parametric ReLU and rational activations \citep{delfosse2020rationals}, no clear winner emerged, and proceeded selecting $tanh$, being the least computationally expensive one.

A major challenge during this phase was the instability in training LSTM models, resulting in frequent \textit{NaN} loss failures despite adjustments to: LR, gradient clipping, loss weights, training samples length, and the use of truncated backpropagation through time (TBPTT) \citep{comunità2025nablafxframeworkdifferentiableblackbox}.
A suboptimal solution was to select the training scheme reported in Table~\ref{tab:training}. 
While this does not prevent \textit{NaN} loss failures, it helps to reach a suitable local optimum by training each model with a broader range of hyperparameters. 
For TBPTT, gradients are updated every 4800 samples (100 ms at 48 kHz), and the max training steps match the epochs used for models without TBPTT.
All other architectures revealed to be much more stable - with no \textit{NaN} loss cases - while still benefiting from training on a variety of LRs to achieve optimal results.
For all architectures we opt to reduce LR by half when there are no improvements in total validation loss for 20 epochs.
To prevent overfitting, we also use early stopping of 50 epochs.

\begin{figure*}[t]
    \begin{minipage}[b]{.32\textwidth}
        \centering
        \includegraphics[width=.9\textwidth]{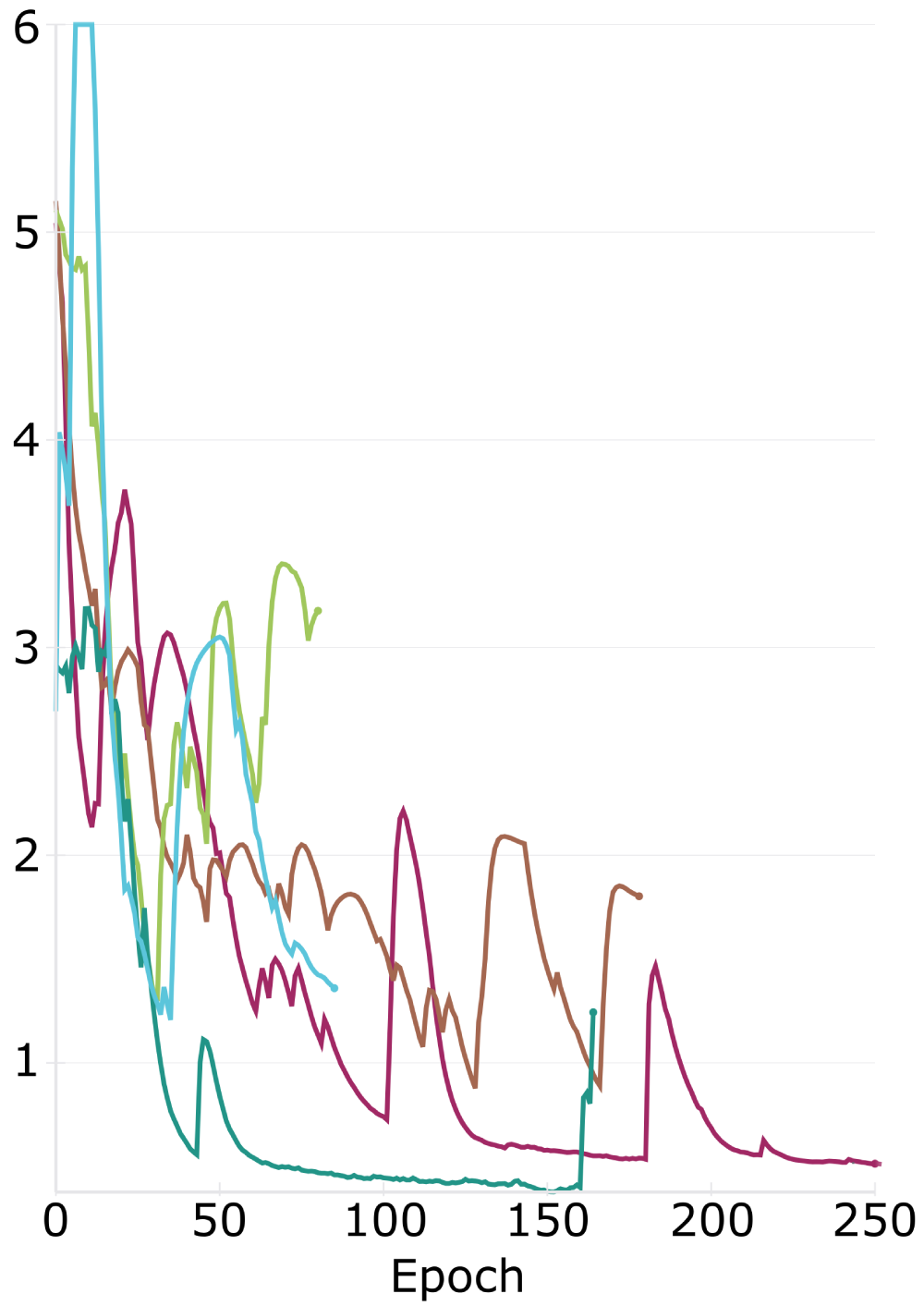}
        \\\textbf{(a)} LSTM trained on\\Harley Benton Plexicon
        \label{fig:ex_loss_lstm}
    \end{minipage}
    \begin{minipage}[b]{.32\textwidth}
        \centering
        \includegraphics[width=.9\textwidth]{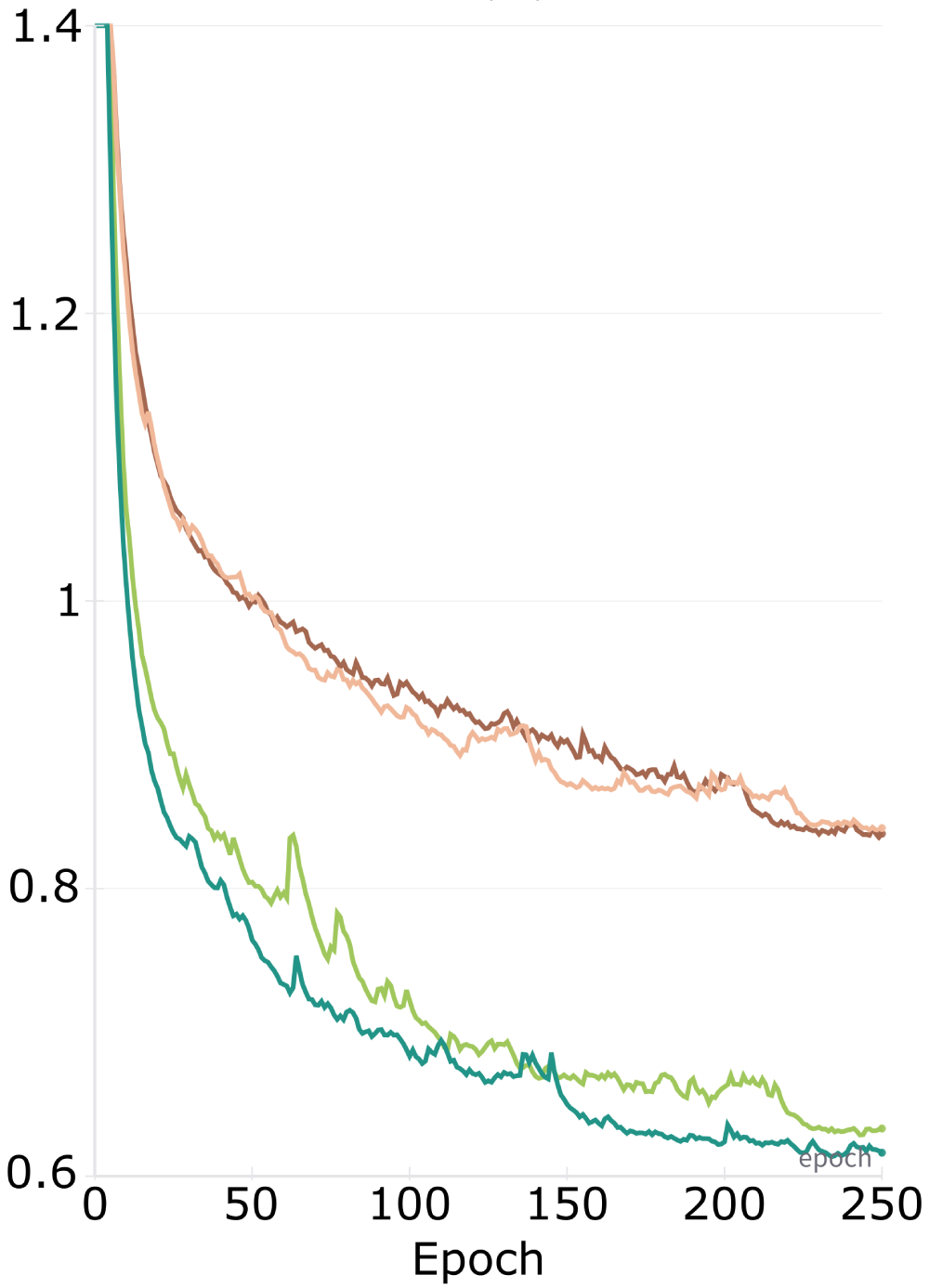}
        \\\textbf{(b)} S4 trained on\\Harley Benton Rodent
        \label{fig:ex_loss_s4}
    \end{minipage}
    \begin{minipage}[b]{.32\textwidth}
        \centering
        \includegraphics[width=.9\textwidth]{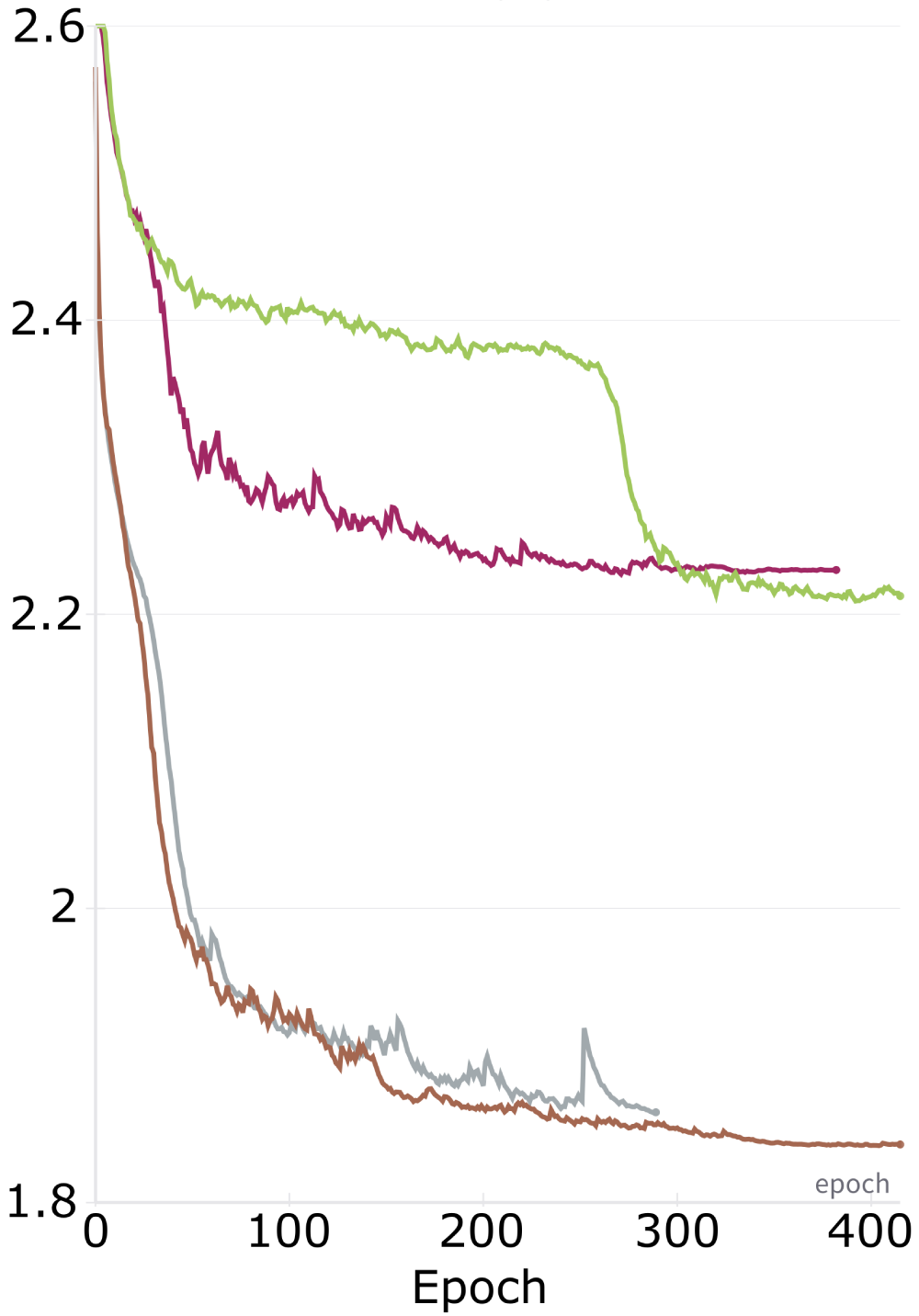}
        \\\textbf{(c)} GB-FUZZ trained on\\Harley Benton Fuzzy Logic
        \label{fig:ex_loss_gb}
    \end{minipage}
    \caption{Validation scaled-loss for different learning rates reveals training discrepancies and unreliability of previous literature comparing architectures and model configurations using a single training scheme.}
    \label{fig:example_loss_variance}
\end{figure*}

Examples that highlight differences in performance as a function of LR are shown in Figure~\ref{fig:example_loss_variance}, with training sessions that might not reach a good local optimum (\ref{fig:example_loss_variance}a) or exhibit scaled-loss differences of up to 30\%-40\% (\ref{fig:example_loss_variance}b and \ref{fig:example_loss_variance}c).

\section{Results}
\label{sec:results}

\subsection{Results for Non-parametric Models}
\label{sec:results_overall}
\begin{figure*}[t]
    \begin{minipage}[b]{1\textwidth}
        \centering
        \includegraphics[height=5cm]{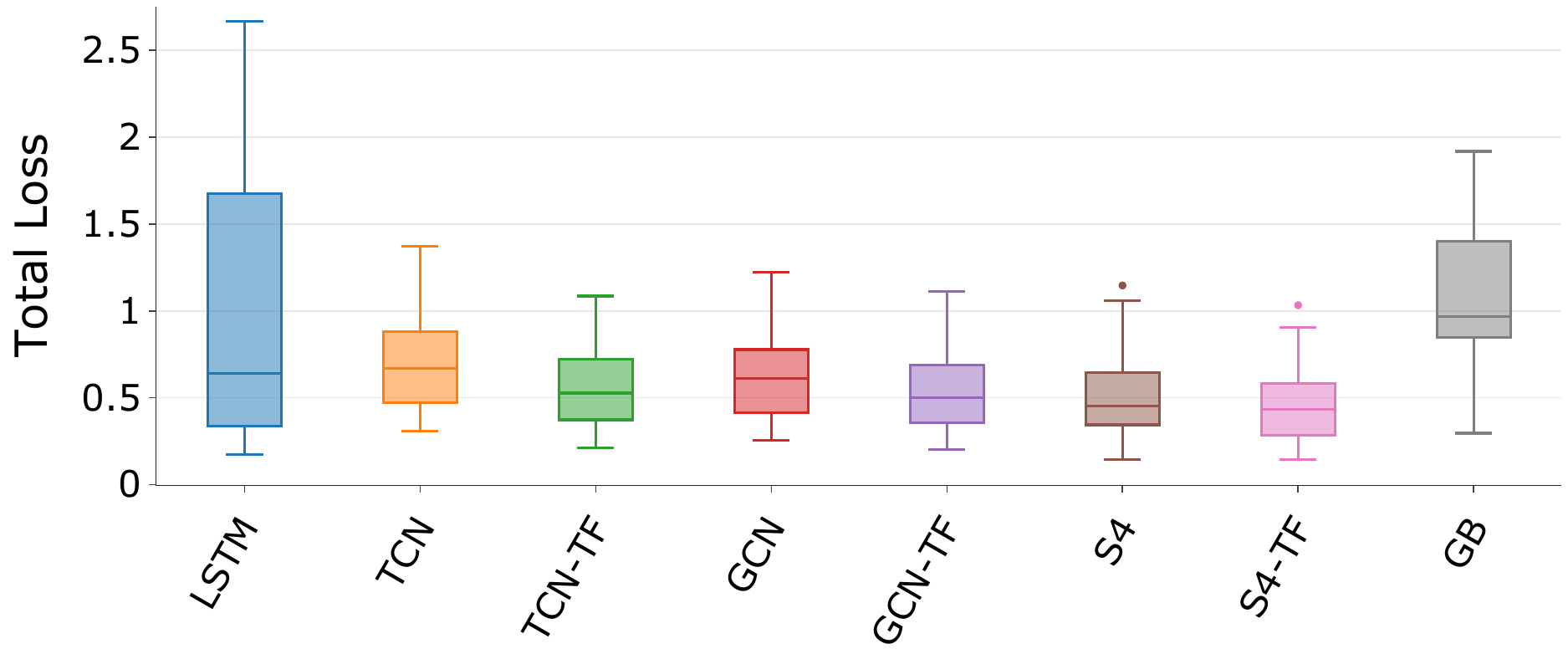}
        \\\textbf{(a)} Overall
        \label{fig:bp_loss_overall}
    \end{minipage}\\\\
    \begin{minipage}[b]{.245\textwidth}
        \centering
        \includegraphics[height=4.0cm]{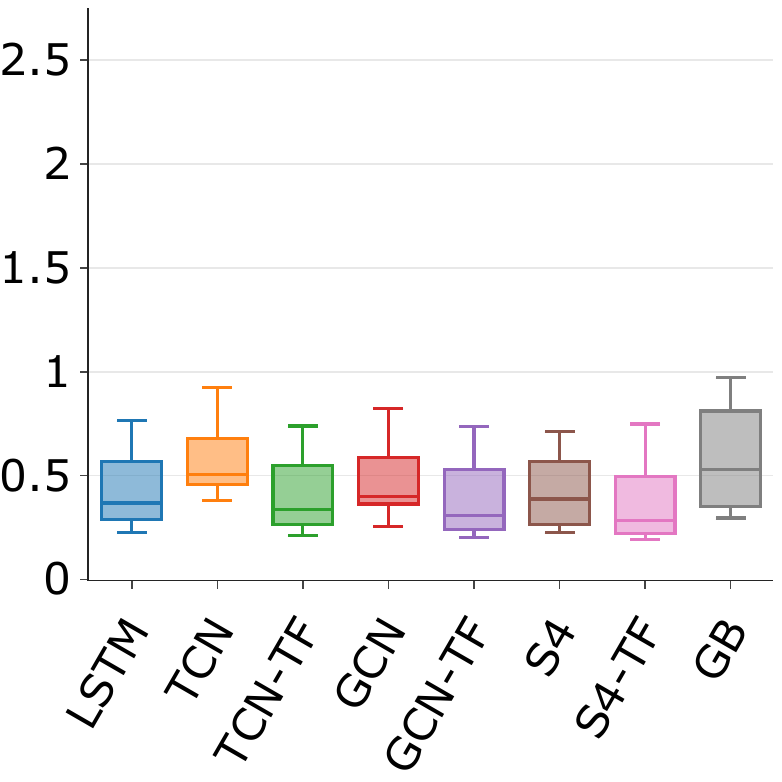}
        \\\textbf{(b)} Compressor/Limiter
        \label{fig:bp_loss_comp}
    \end{minipage}
    \begin{minipage}[b]{.245\textwidth}
        \centering
        \includegraphics[height=4.0cm]{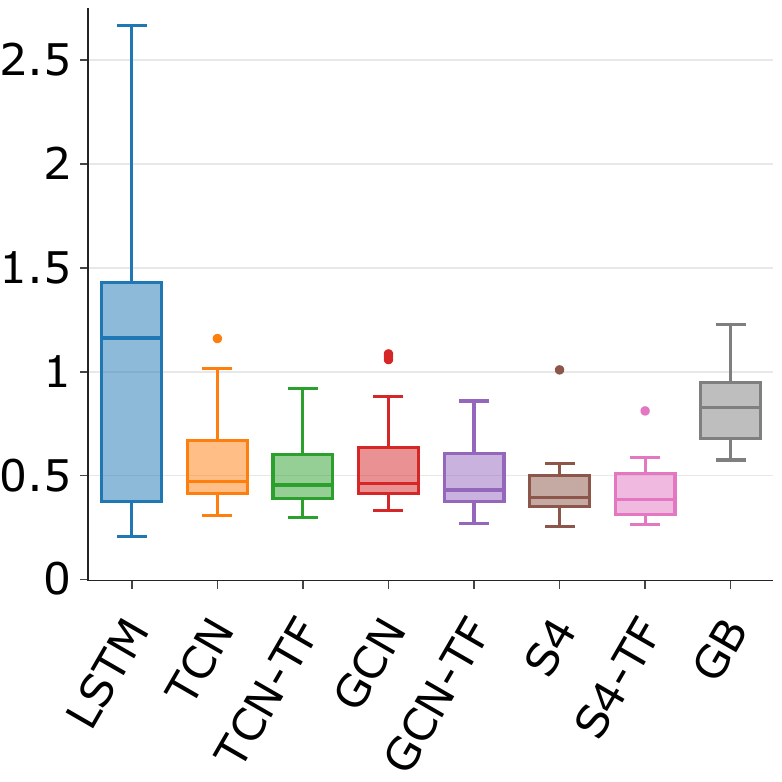}
        \\\textbf{(c)} Overdrive
        \label{fig:bp_loss_od}
    \end{minipage}
    \begin{minipage}[b]{.245\textwidth}
        \centering
        \includegraphics[height=4.0cm]{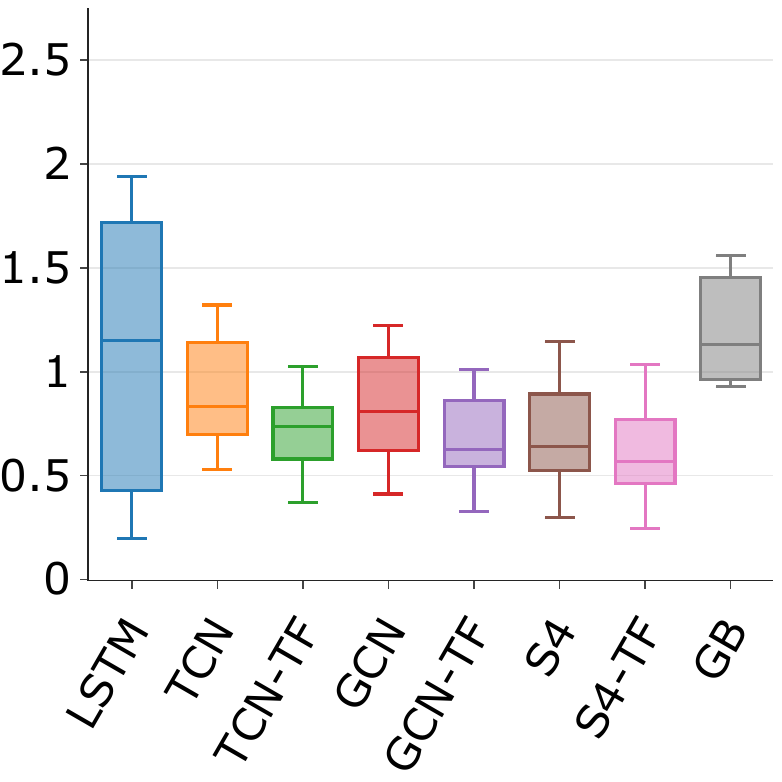}
        \\\textbf{(d)} Distortion
        \label{fig:bp_loss_dist}
    \end{minipage}
    \begin{minipage}[b]{.245\textwidth}
        \centering
        \includegraphics[height=4.0cm]{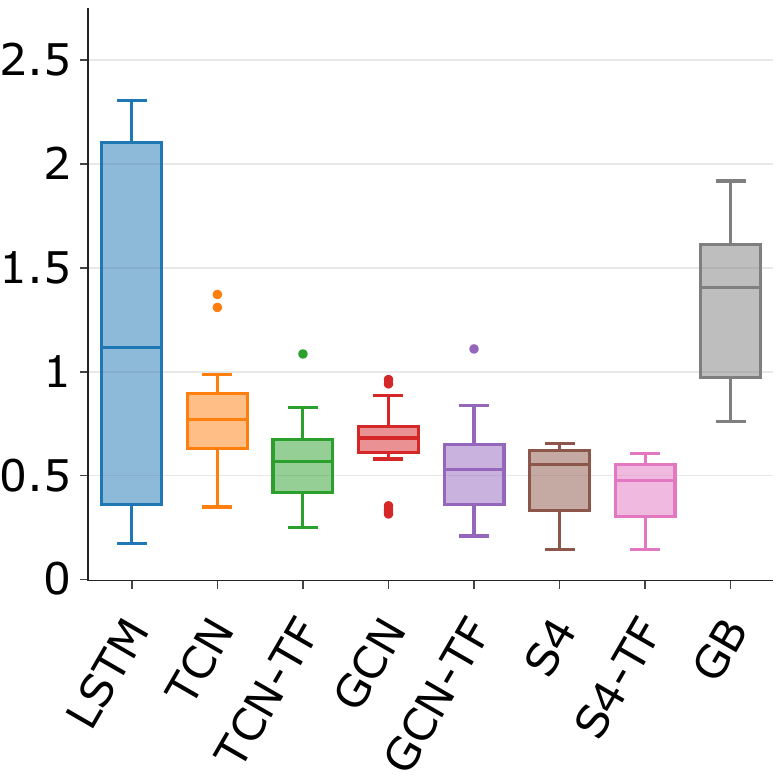}
        \\\textbf{(e)} Fuzz
        \label{fig:bp_loss_fuzz}
    \end{minipage}
    \caption{Total loss for different architectures: overall and for each effect type}
    \label{fig:bp_loss}
\end{figure*}

Figure~\ref{fig:bp_loss} shows the total loss for architectures trained on sixteen effects (see Sec.~\ref{tab:effects_configs}), with box plots for overall results (\ref{fig:bp_loss}a) and each effect category (\ref{fig:bp_loss}b-e).

Overall, S4 models with TFiLM conditioning perform the best.
Also, TFiLM improves median performance across architectures and reduces the variance, demonstrating how time-varying conditioning enhances expressivity, regardless of the backbone.
Although GCN achieves lower loss than TCN, TFiLM conditioning results in similar performance for both, making the added complexity of GCN-TF hard to justify and emphasizing how efficient and effective time-varying conditioning could be focus of further research.
Gray-box models show higher median loss than black-box ones but can outperform them in some cases, and also exhibit smaller variance than LSTMs.
Having the greatest variance across architectures, LSTMs are shown to be the least reliable, which makes them unsuitable for modeling a wide range of effect types and devices.

Breaking down performance by effect type, S4 models perform best in each category, with TFiLM conditioning generally improving or maintaining performance without hindering it.
It therefore seems that S4 models with a form of time-dependent conditioning are the best candidate to develop high performing architectures across a wide range of effect types and implementations.

In convolutional architectures, TFiLM is shown to be helpful for every effect category although to a varying extent, particularly benefiting distortion and fuzz effects with their complex timbre and time-dependent behaviors.
This is reflected in the average loss and standard deviation in Table~\ref{tab:results_overall}, where distortion and fuzz are the most challenging effects to model.
Beside compressors, LSTM is confirmed to be the least reliable architecture in each category, with far greater variance than any other architecture and performing on par or worse than gray-box models.

Looking in further details at the results in Table~\ref{tab:results_overall} where we gather mean ($\mu$) and standard deviation ($\sigma$) of the total loss for each device type and overall, we notice how S4 with TFiLM conditioning (S4-TF-L-16) performed the best, with TFiLM giving only a marginal improvement w.r.t. base S4 models of the same \textit{size}.
Standard S4 and GCN-TF rank second and third in average loss, while TCN-TF has the lowest standard deviation across effect types.
TFiLM conditioning consistently improves loss, especially for convolutional backbones, enhancing performance at each receptive field and model size, while enabling smaller models to match the performance of larger ones, reducing the reliance on receptive field length and nonlinear activations to capture complex timbres and long-range dependencies.
Overall, TCN-TF models perform well in terms of mean loss and are among the best in terms of standard deviation, which makes them a good choice to model diverse effects.
Once more, looking at each model size and receptive field length, the performance improvement of GCN and GCN-TF w.r.t. TCN and TCN-TF models does not seem to justify the increase in parameters and computational cost.
Also worth noticing is that rational nonlinearities are useful for gray-box models, allowing to match performance while greatly reducing the computational complexity of MLP-based models, regardless of the effect and across architectures (i.e., GB-DIST and GB-FUZZ).

\setlength{\tabcolsep}{4pt}
\renewcommand{\arraystretch}{1.3}
\begin{table*}[ht!]
    \caption{Mean ($\mu$) and Standard Deviation ($\sigma$) of the total loss (L1 + MR-STFT) - overall and for each device type - for non parametric models. Bold indicates top performing model. Underline indicates bottom performing model. Superscripts indicate top-3 performing models. Subscripts indicate bottom-3 performing models.}
    \label{tab:results_overall}
    \centerline{
        \begin{tabular}{lccc|cc|cc|cc|cc} 
            \midrule
            \midrule
            
            \multirow{2}{*}{Model} 
                & \multirow{2}{*}{\# Params.} 
                        & \multicolumn{2}{c}{Comp./Limit.} 
                            &  \multicolumn{2}{c}{Overdrive} 
                                & \multicolumn{2}{c}{Distortion} 
                                    & \multicolumn{2}{c}{Fuzz} 
                                        & \multicolumn{2}{c}{Overall} \\
            \cmidrule(lr){3-4} 
                \cmidrule(lr){5-6} 
                    \cmidrule(lr){7-8} 
                        \cmidrule(lr){9-10}
                            \cmidrule(lr){11-12}
            &   & $\mu$ & $\sigma$ & $\mu$ & $\sigma$ & $\mu$ & $\sigma$ & $\mu$ & $\sigma$ & $\mu$ & $\sigma$\\
            \hline
            LSTM-32
                & 4.5k  & 0.4342 & 0.2120 & \underline{1.3099} & \underline{1.0003} & 1.0956 & 0.6605 & 1.2560 & 0.9954 & 1.0239 & \underline{0.7904}$_{[1]}$\\ 
            LSTM-96       
                & 38.1k & 0.4276 & 0.2353 & 0.8932 & 0.6595 & 1.0873 & \underline{0.8546} & 1.1554 & \underline{1.0093} & 0.8909 & \underline{0.7308}$_{[2]}$\\ 
            \hline
            TCN-45-S-16               
                & 7.5k & 0.5869	& 0.2363 & 0.5951 & 0.2727 & 0.9274 & 0.2865 & 0.9233 & 0.3039 & 0.7582 & 0.3012\\ 
            TCN-45-L-16               
                & 7.3k & \underline{0.5948}	& 0.1994 & 0.5490 & 0.2383 & 0.9708	& 0.2210 & 0.7839 & 0.1655 & 0.7246 & 0.2537\\
            TCN-250-S-16               
                & 14.5k & 0.5869 & 0.1820 & 0.5847 & 0.2893 & 0.8565 & 0.3099 & 0.7335 & 0.2515 & 0.6904 & 0.2625\\ 
            TCN-250-L-16               
                & 18.4k & 0.5560 & 0.1830 & 0.5048 & 0.2196 & 0.8594 & 0.2764 & 0.6640 & 0.1893 & 0.6461 & 0.2418\\
            TCN-2500-S-16               
                & 13.7k & 0.5803 & 0.1695 & 0.6516 & 0.3396 & 0.9248 & 0.2849 & 0.8723 & 0.3959 & 0.7573 & \underline{0.3142}$_{[3]}$\\ 
            TCN-2500-L-16               
                & 11.9k & 0.5670 & \textbf{0.1533} & 0.5268	& 0.2314 & 0.8162 & 0.2914 & 0.6460	& 0.2202 & 0.6390 & 0.2350\\
            \hline
            TCN-TF-45-S-16               
                & 39.5k & 0.4010 & 0.2161 & 0.5416 & 0.1907 & 0.7701 & 0.1957 & 0.6045 & 0.2072 & 0.5793 & 0.2272\\
            TCN-TF-45-L-16               
                & 71.3k & 0.4024 & 0.2272 & 0.4852 & 0.1394 & 0.7083 & 0.1877 & 0.4990 & 0.1659 & 0.5237 & \textbf{0.2008}$^{[1]}$\\
            TCN-TF-250-S-16               
                & 52.9k & 0.4085 & 0.2281 & 0.4909 & 0.1793 & 0.7130 & 0.2154 & 0.5435 & 0.1582 & 0.5390 & 0.2106\\
            TCN-TF-250-L-16               
                & 88.8k & 0.3973 & 0.2066 & 0.4631 & 0.1704 & 0.6665 & 0.2191 & 0.5120 & 0.1832 & 0.5097 & \textbf{0.2029}$^{[2]}$\\
            TCN-TF-2500-S-16               
                & 45.7k & 0.4275 & 0.2103 & 0.5704 & 0.2339 & 0.7524 & \textbf{0.1508} & 0.6600	& 0.3208 & 0.6026 & 0.2454\\
            TCN-TF-2500-L-16               
                & 75.9k & 0.4139 & 0.2116 & 0.5077 & 0.2430 & 0.6422 & 0.1813 & 0.5588 & 0.2390 & 0.5307 & 0.2146\\
            \hline
            GCN-45-S-16               
                & 16.2k & 0.5225 & 0.2011 & 0.6151 & 0.3170 & 0.8718 & 0.3182 & 0.7694 & 0.1622 & 0.6947 & 0.2705\\ 
            GCN-45-L-16               
                & 17.1k & 0.5195 & 0.1917 & 0.5380 & 0.2388 & 0.8502 & 0.3146 & 0.7237 & \textbf{0.1533} & 0.6579 & 0.2514\\
            GCN-250-S-16               
                & 30.4k & 0.4713 & 0.1942 & 0.5956 & 0.3101 & 0.8138 & 0.3170 & 0.6350 & 0.2072 & 0.6289 & 0.2674\\ 
            GCN-250-L-16               
                & 39.6k & 0.4747 & 0.1899 & 0.4896 & 0.1955 & 0.7838 & 0.3312 & 0.6320 & 0.2243 & 0.5950 & 0.2521\\
            GCN-2500-S-16               
                & 28.6k & 0.4451 & 0.2069 & 0.6179 & 0.3047 & 0.8475 & 0.2696 & 0.5986 & 0.1842 & 0.6273 & 0.2655\\ 
            GCN-2500-L-16               
                & 26.4k & 0.4274 & 0.2022 & 0.4909 & 0.1778 & 0.7734 & 0.2870 & 0.6112 & 0.1980 & 0.5757 & 0.2395\\
            \hline
            GCN-TF-45-S-16               
                & 141.6k & 0.3787 & 0.2319 & 0.5288 & 0.1814 & 0.7110 & 0.2228 & 0.5386	& 0.2208 & 0.5393 & 0.2276\\
            GCN-TF-45-L-16               
                & 268.0k & 0.3783 & 0.2192 & 0.4632 & 0.1573 & 0.6605 & 0.2377 & 0.4452	& 0.1595 & 0.4868 & 0.2067\\
            GCN-TF-250-S-16               
                & 181.0k & 0.3882 & 0.2263 & 0.4789 & 0.1352 & 0.7005 & 0.2523 & 0.6323	& 0.3734 & 0.5500 & 0.2659\\
            GCN-TF-250-L-16               
                & 315.6k & 0.3857 & 0.2301 & 0.4436	& 0.1563 & 0.6579 & 0.2604 & 0.5564	& 0.2689 & 0.5109 & 0.2349\\
            GCN-TF-2500-S-16               
                & 154.1k & 0.4006 & 0.2290 & 0.5290 & 0.2227 & 0.7034 & 0.2127 & 0.5209	& 0.1859 & 0.5385 & 0.2209\\
            GCN-TF-2500-L-16               
                & 277.3k & 0.3934 & 0.2335 & 0.4496	& 0.1627 & 0.6493 & 0.2362 & 0.4547	& 0.1667 & \textbf{0.4867}$^{[3]}$ & 0.2072\\
            \hline
            S4-S-16              
                & 2.4k & 0.4623	& 0.1893 & 0.5441	& 0.3122 & 0.7422 & 0.2931 & 0.5309	& 0.2030 & 0.5699 & 0.2523\\
            S4-L-16              
                & 19.0k & 0.3839 & 0.2035 & \textbf{0.3964} & \textbf{0.1300} & 0.6490 & 0.3136 & 0.4217 & 0.1949 & \textbf{0.4627}$^{[2]}$ & 0.2269\\
            \hline
            S4-TF-S-16              
                & 28.0k & 0.3933 & 0.2422 & 0.4726 & 0.2321 & 0.6588 & 0.2655 & 0.4325 & 0.1798 & 0.4893 & 0.2326\\
            S4-TF-L-16             
                & 70.2k & \textbf{0.3438} & 0.2119 & 0.4001	& 0.1422 & \textbf{0.5599} & 0.2718 & \textbf{0.4210} & 0.1958 & \textbf{0.4312}$^{[1]}$ & \textbf{0.2054}$^{[3]}$\\
            \hline
            GB-COMP
                & 47 & 0.5823 & \underline{0.2987} & - & - & - & - & - & - & 0.5823	& 0.2987\\
            \hline
            GB-DIST-MLP             
                & 2.2k & - & - & 0.8390	& 0.2840 & 1.1898 & 0.2909 & 1.4419	& 0.4511 & \underline{1.1569}$_{[2]}$ & 0.4090\\
            GB-DIST-RNL             
                & 47 & - & - & 0.8291 & 0.1482 & 1.1938 & 0.2956 & \underline{1.5086} & 0.4901 & \underline{1.1771}$_{[1]}$ & 0.4236\\
            \hline
            GB-FUZZ-MLP             
                & 2.3k & - & - & 0.8500	& 0.2238 & 1.1920 & 0.2853 & 1.1962	& 0.3108 & 1.0794 & 0.3015\\
            GB-FUZZ-RNL             
                & 62 & - & - & 0.8636 & 0.2542 & \underline{1.2135}	& 0.2823 & 1.2288 & 0.3109 & \underline{1.1020}$_{[3]}$ & 0.3110\\
            \hline
            Average 
                &   & 0.4560 & 0.2120 & 0.5914 & 0.2557 & 0.8298 & 0.2910 & 0.7268 & 0.2837\\
            \hline
            \hline
        \end{tabular}
    }
\end{table*}

\setlength{\tabcolsep}{3pt}
\renewcommand{\arraystretch}{1.3}
\begin{table*}[ht!]
    \caption{Mean ($\mu$) and Standard Deviation ($\sigma$) of the total loss (L1 + MR-STFT) - overall and for each device type - for non parametric models. Bold indicates top performing model. Underline indicates bottom performing model. Superscripts indicate top-3 performing models. Subscripts indicate bottom-3 performing models.}
    \label{tab:results_overall_param}
    \centerline{
        \begin{tabular}{l c|cc >{\columncolor{gray!20}}c|cc c|cc >{\columncolor{gray!20}}c|cc}
            \midrule
            \midrule
            \multirow{4}{*}{Model}
            & \multicolumn{6}{c}{Marshall JVM410H}
            & \multicolumn{6}{c}{Multidrive Pedal Pro F-Fuzz}\\
            
            \cmidrule(lr){2-7} 
            \cmidrule(lr){8-13} 

            & \multicolumn{3}{c}{Val. Loss}
            & \multicolumn{3}{c}{Test Loss}
            & \multicolumn{3}{c}{Val. Loss}
            & \multicolumn{3}{c}{Test Loss}\\
        
            \cmidrule(lr){2-4}
            \cmidrule(lr){5-7}
            \cmidrule(lr){8-10}
            \cmidrule(lr){11-13}
            
            & Tot. & {\footnotesize $L1$} & {\footnotesize MR-STFT} 
            & Tot. & {\footnotesize $L1$} & {\footnotesize MR-STFT} 
            & Tot. & {\footnotesize $L1$} & {\footnotesize MR-STFT} 
            & Tot. & {\footnotesize $L1$} & {\footnotesize MR-STFT}\\ 
            
            \hline
            LSTM-C-32 & 0.4634 & 0.0131 & 0.4503 & 1.1216 & 0.0376 & 1.0839 & 0.3234 & 0.0050 & 0.3184 & 0.3123 & 0.0051 & 0.3072\\
            LSTM-TVC-32 & 0.4521 & 0.0103 & 0.4419 & 1.1610 & 0.0351 & 1.1259 & 0.2071 & 0.0027 & 0.2044 & \textbf{0.1652}$^{[2]}$ & 0.0023 & 0.1628\\
            \hline
            LSTM-C-96 & 0.4686 & 0.0126 & 0.4560 & 1.1423 & 0.0347 & 1.1075 & 0.2405 & 0.0052 & 0.2353 & \textbf{0.1689}$^{[3]}$ & 0.0024 & 0.1665\\
            LSTM-TVC-96 & 0.4345 & 0.0098 & 0.4247 & 1.1142 & 0.0344 & 1.0798 & 0.2026 & 0.0053 & 0.1973 & \textbf{0.1560}$^{[1]}$ & 0.0040 & 0.1521\\
            \hline
            \hline
            TCN-F-45-S-16 & 0.6137 & 0.0224 & 0.5913 & 1.2636 & 0.0519 & 1.2117 & 0.6664 & 0.0163 & 0.6501 & 0.7095 & 0.0217 & 0.6878\\
            TCN-TF-45-S-16 & 0.4624 & 0.0115 & 0.4509 & 1.2087 & 0.0333 & 1.1754 & 0.5147 & 0.0082 & 0.5065 & 0.4886 & 0.0077 & 0.4809\\
            TCN-TTF-45-S-16 & 0.5220 & 0.0163 & 0.5057 & 1.2982 & 0.0449 & 1.2533 & 0.5594 & 0.0106 & 0.5488 & 0.5324 & 0.0102 & 0.5223\\
            TCN-TVF-45-S-16 & 0.4809 & 0.0164 & 0.4645 & 1.4137 & 0.0534 & 1.3603 & 0.5491 & 0.0117 & 0.5374 & 0.5356 & 0.0115 & 0.5241\\
            \hline
            TCN-F-45-L-16 & 0.5962 & 0.0230 & 0.5733 & \underline{1.5938}$_{[1]}$ & 0.0638 & 1.5300 & 0.6496 & 0.0176 & 0.6320 & 0.6681 & 0.0185 & 0.6495\\
            TCN-TF-45-L-16 & 0.3976 & 0.0090 & 0.3886 & 1.1138 & 0.0319 & 1.0819 & 0.4026 & 0.0062 & 0.3964 & 0.3553 & 0.0058 & 0.3495\\
            TCN-TTF-45-L-16 & 0.4759 & 0.0134 & 0.4626 & 1.2022 & 0.0385 & 1.1637 & 0.5052 & 0.0263 & 0.4790 & 0.4788 & 0.0225 & 0.4563\\
            TCN-TVF-45-L-16 & 0.4476 & 0.0134 & 0.4343 & 1.1835 & 0.0393 & 1.1442 & 0.5872 & 0.0171 & 0.5701 & 0.5835 & 0.0164 & 0.5671\\
            \hline
            \hline
            S4-F-S-16 & 0.4571 & 0.0089 & 0.4482 & 1.2783 & 0.0465 & 1.2317 & 0.5171 & 0.0107 & 0.5064 & 0.7687 & 0.0243 & 0.7444\\
            S4-TF-S-16 & 0.3864 & 0.0073 & 0.3791 & \textbf{1.0965}$^{[2]}$ & 0.0290 & 1.0675 & 0.3710 & 0.0055 & 0.3655 & 0.4034 & 0.0075 & 0.3959\\
            S4-TTF-S-16 & 0.4227 & 0.0102 & 0.4125 & 1.2164 & 0.0368 & 1.1796 & 0.4264 & 0.0066 & 0.4198 & 0.3816 & 0.0066 & 0.3749\\
            S4-TVF-S-16 & 0.3778 & 0.0095 & 0.3683 & 1.0991	& 0.0351 & 1.0640 & 0.3673 & 0.0055 & 0.3618 & 0.3354 & 0.0058 & 0.3296\\
            \hline
            S4-F-L-16 & 0.3503 & 0.0084 & 0.3419 & 1.1745 & 0.0374 & 1.1370 & 0.3811 & 0.0262 & 0.3549 & 0.4973 & 0.0225 & 0.4748\\
            S4-TF-L-16 & 0.3109 & 0.0049 & 0.3060 & 1.1490 & 0.0321 & 1.1169 & 0.2907 & 0.0054 & 0.2853 & 0.2619 & 0.0042 & 0.2577\\
            S4-TTF-L-16 & 0.4066 & 0.0132 & 0.3933 & \textbf{1.0984}$^{[3]}$ & 0.0349 & 1.0635 & 0.3506 & 0.0071 & 0.3436 & 0.3683 & 0.0075 & 0.3608\\
            S4-TVF-L-16 & 0.2965 & 0.0048 & 0.2917 & \textbf{1.0133}$^{[1]}$ & 0.0257 & 0.9876 & 0.2476 & 0.0041 & 0.2435 & 0.2673 & 0.0045 & 0.2628\\
            \hline
            \hline
            GB-C-DIST-MLP & 0.8563 & 0.0413 & 0.8151 & 1.5072 & 0.0741 & 1.4331 & 1.1759 & 0.0631 & 1.1128 & \underline{1.2104}$_{[2]}$ & 0.0611 & 1.1492\\
            GB-C-DIST-RNL & 0.8395 & 0.0407 & 0.7988 & \underline{1.5113}$_{[3]}$ & 0.0667 & 1.4445 & 1.2355 & 0.0683 & 1.1671 & \underline{1.2531}$_{[1]}$ & 0.0672 & 1.1858\\
            \hline
            GB-C-FUZZ-MLP & 0.8208 & 0.0401 & 0.7807 & 1.5009 & 0.0686 & 1.4323 & 0.9809 & 0.0363 & 0.9446 & 0.9303 & 0.0345 & 0.8958\\
            GB-C-FUZZ-RNL & 0.8123 & 0.0438 & 0.7685 & \underline{1.5428}$_{[2]}$ & 0.0720 & 1.4708 & 1.0043 & 0.0398 & 0.9645 & \underline{0.9395}$_{[3]}$ & 0.0355 & 0.9040\\
            \hline
            \hline
        \end{tabular}
    }
\end{table*}

\subsection{Results for Parametric Models}
\label{sec:results_parametric_models}

Table~\ref{tab:results_overall_param} shows the losses for parametric models trained on Marshall JVM410H guitar amplifier and Multidrive Pedal Pro F-Fuzz digital emulation of Fuzz Face guitar pedal.
We report both validation and test loss to assess model performance when tested on unseen controls configurations (i.e., guitar amp) or unseen sources and content (i.e., fuzz pedal).
S4 models are the best performing on the guitar amp, with LSTM models and TCN with time-varying conditioning performing only marginally worse.
For the fuzz effect pedal emulation, LSTMs are the best performing, with S4 models with time-varying conditioning (S4-TF-L-16 and S4-TVF-L-16) being only slightly worse.

While TVCond (TVC) conditioning is not useful to improve LSTM performance on the Marshall amp, it is beneficial to achieve among best performance with a smaller LSTM model.
For TCN models, TFiLM (TF) conditioning is the most effective for both guitar amp and fuzz pedal cases, regardless of the number of layers in the model.
We also notice how the baseline FiLM (F) conditioning is the worst mechanism for all but one case when applied to either TCN or S4 models.
Looking at S4 models, the proposed TTFiLM (TTF) and TVFiLM (TVF) conditioning allow performance on par or better than TFiLM (TF) regardless of model size.
Also, TVF allows small models to achieve performance comparable to bigger ones.
For TCN models, it is less clear whether there is an advantage in using TTF or TVF w.r.t. TF beside computational complexity.

To recap, TFiLM seems to be the best across architectures, with TVFiLM better than TTFiLM when reducing the computational complexity.
A study on a wider variety of effects would allow to better asses whether TVF and TTF can compete with TF conditioning.

\begin{figure*}[ht!]
    \begin{center}
    \begin{minipage}[b]{.32\textwidth}
        \centering
        \includegraphics[height=4.5cm]{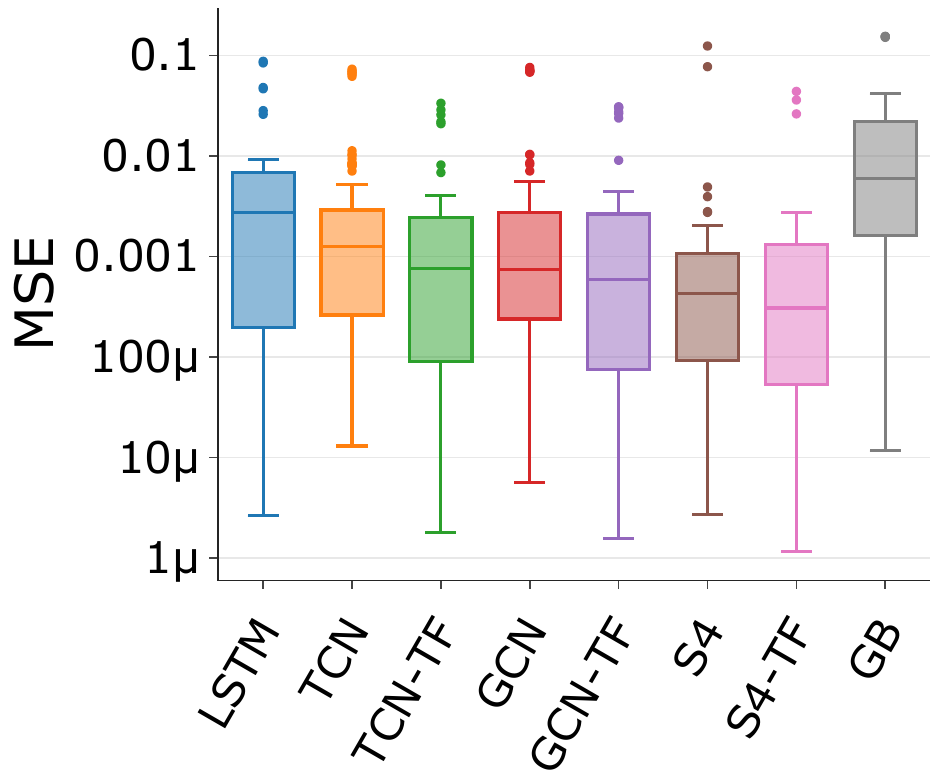}
        \\\textbf{(a)} MSE
        \label{fig:ex_loss_lstm}
    \end{minipage}
    \begin{minipage}[b]{.32\textwidth}
        \centering
        \includegraphics[height=4.5cm]{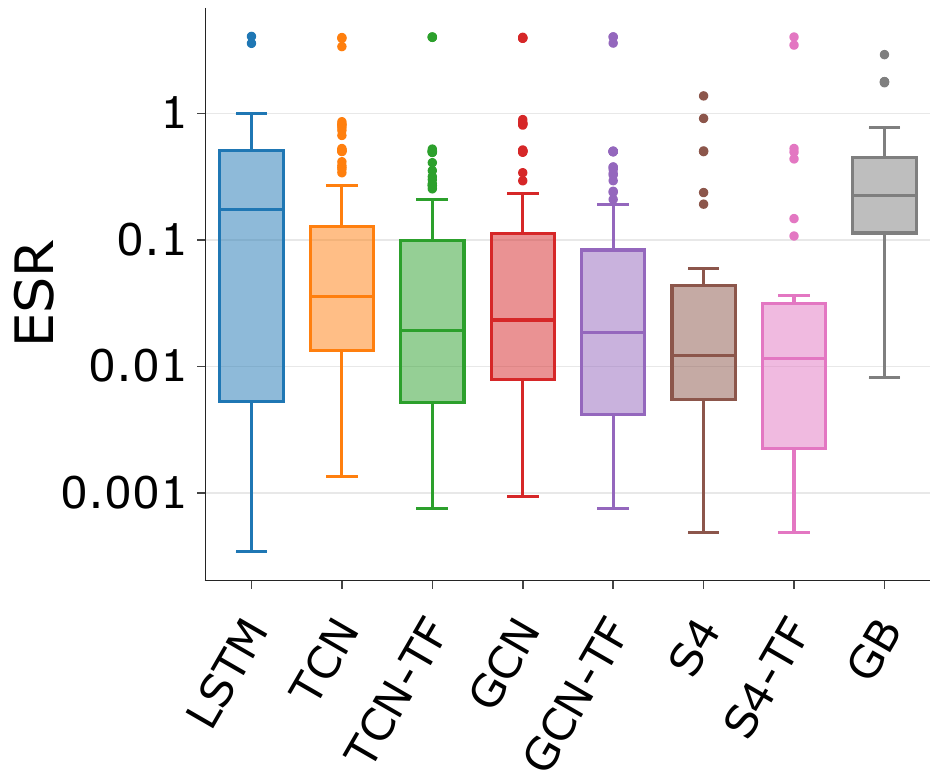}
        \\\textbf{(b)} ESR
        \label{fig:ex_loss_s4}
    \end{minipage}
    \begin{minipage}[b]{.32\textwidth}
        \centering
        \includegraphics[height=4.5cm]{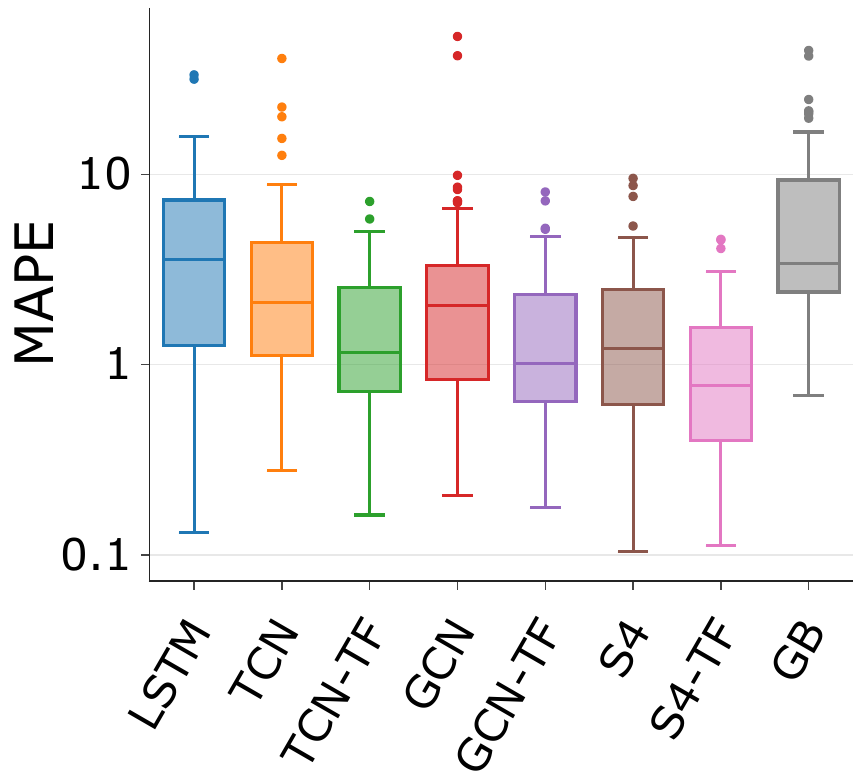}
        \\\textbf{(c)} MAPE
        \label{fig:ex_loss_s4}
    \end{minipage}
    \caption{Metrics for different architectures for all effect types}
    \label{fig:metrics_boxplots}
    \end{center}
\end{figure*}

\begin{figure*}[ht!]
    \centering
    \begin{minipage}[b]{.45\textwidth}
        \centering
        \includegraphics[height=4.8cm]{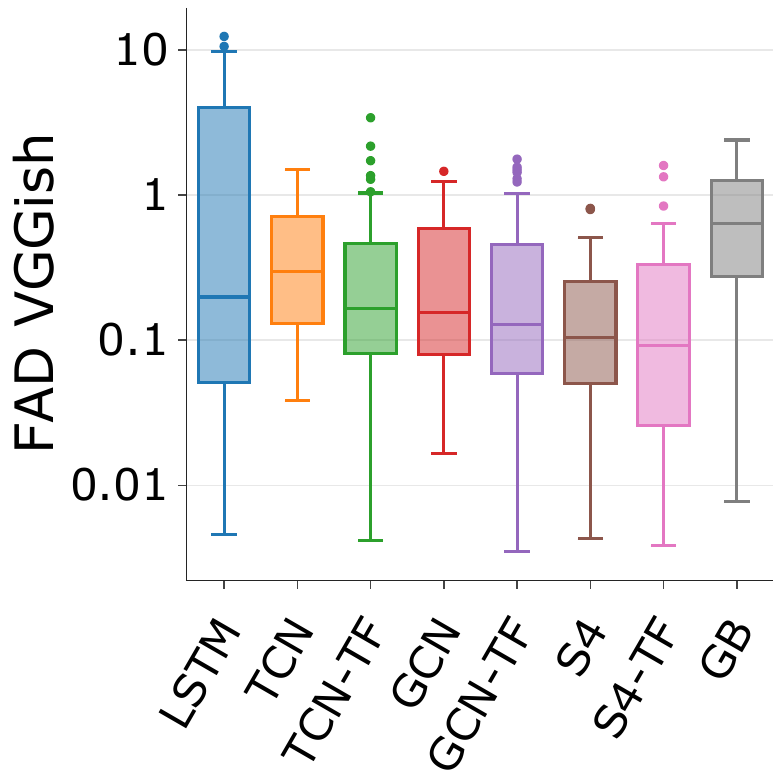}
        \\\textbf{(a)} VGGish
        \label{fig:ex_loss_lstm}
    \end{minipage}
    \begin{minipage}[b]{.45\textwidth}
        \centering
        \includegraphics[height=4.8cm]{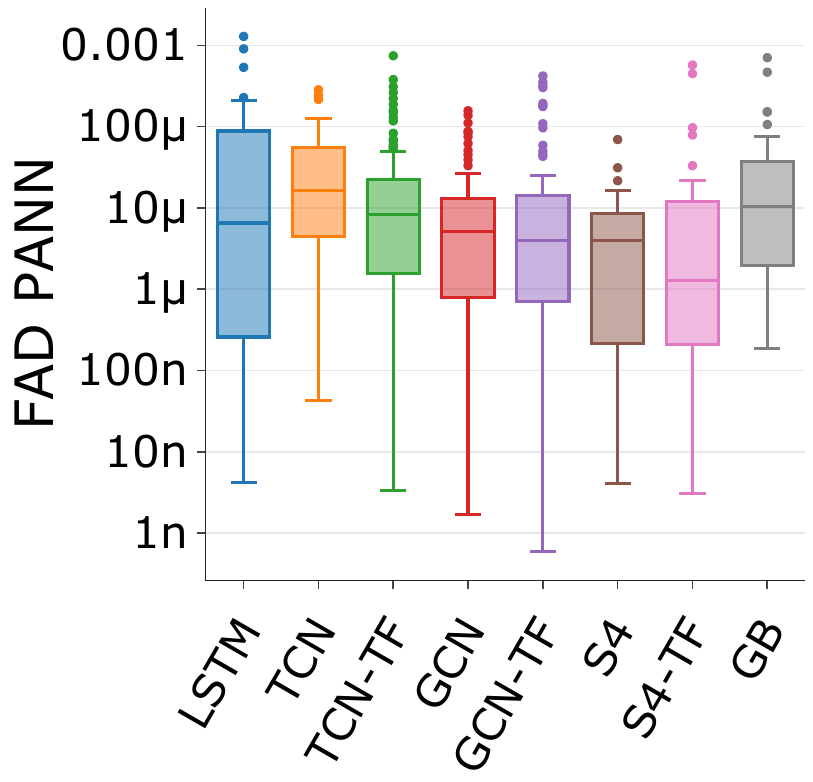}
        \\\textbf{(b)} PANN
        \label{fig:ex_loss_s4}
    \end{minipage}
    
    \begin{minipage}[b]{.45\textwidth}
        \centering
        \includegraphics[height=4.8cm]{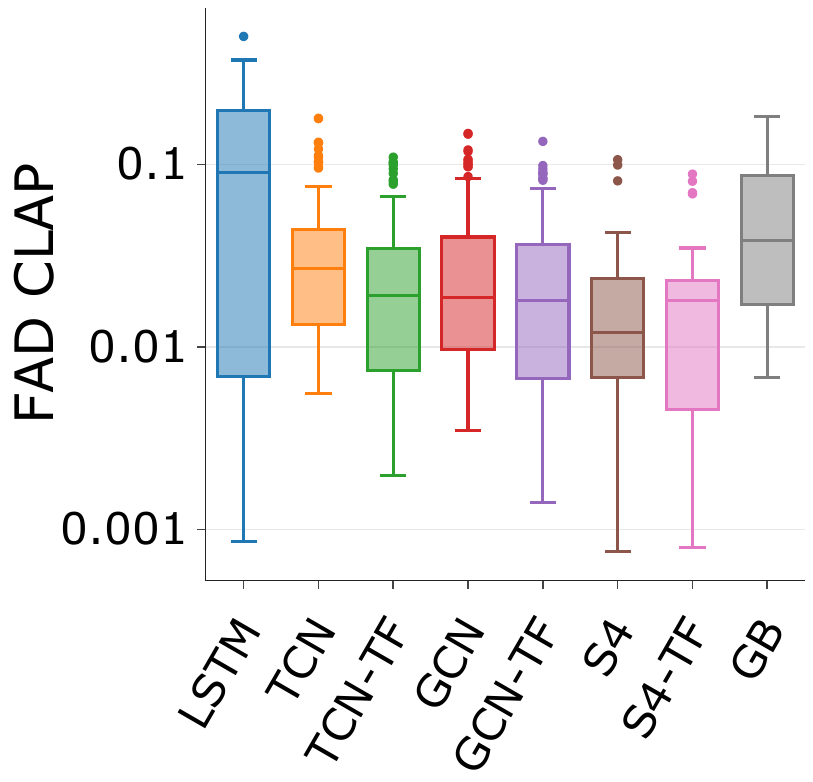}
        \\\textbf{(c)} CLAP
        \label{fig:ex_loss_s4}
    \end{minipage}
    \begin{minipage}[b]{.45\textwidth}
        \centering
        \includegraphics[height=4.8cm]{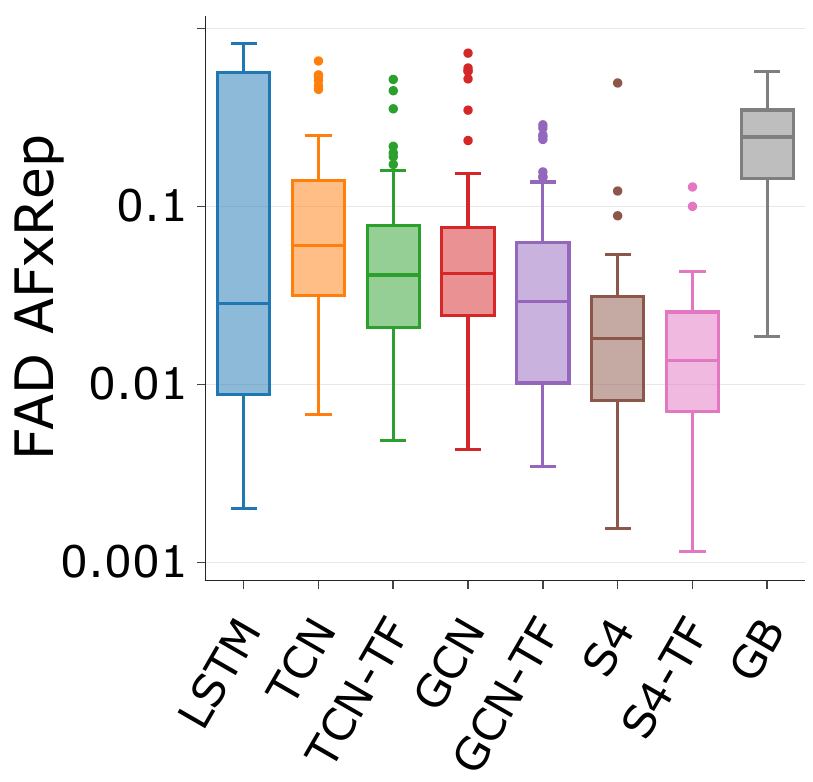}
        \\\textbf{(d)} AFx-Rep
        \label{fig:ex_loss_s4}
    \end{minipage}
    \caption{FAD for different architectures for all effect types}
    \label{fig:metrics_boxplots_fad}
\end{figure*}

\section{Evaluation}
\label{sec:eval}

\subsection{Objective Evaluation}
\label{sec:obj_eval}

To extend the objective evaluation beyond loss values we use most of the metrics available from the NablAFx framework \citep{comunità2025nablafxframeworkdifferentiableblackbox}, these include both signal-based metrics (i.e., MSE, ESR, MAPE) as well as latent representation-based ones (i.e., FAD). 
FAD is computed using three widely adopted representations (i.e., VGGish, PANN and CLAP) as well as a recently proposed audio production style representation (AFx-Rep) \cite{steinmetz2024st}.
We compute these metrics for several reasons beside broadening the analysis: to investigate which ones might be most suited for modeling tasks, to investigate correlation between objective metrics and subjective ratings (see Sec.~\ref{sec:subj_eval}), to encourage further research on audio effects related representations and evaluation methods.

When using objective metrics - which we plot in Fig.~\ref{fig:metrics_boxplots} and \ref{fig:metrics_boxplots_fad} on a log scale to highlight differences - the picture is similar to the results discussed in previous sections.
S4 and S4-TF architectures show the best performance across effect type, regardless of the metric used.
Once again, temporal conditioning (TFiLM), is helpful in improving performance - noticeable from the reduction in median as well as standard deviation values - regardless of the architecture's backbone (TCN, GCN or S4).
LSTM architectures are shown again to have a high performance variance, making them not reliable when modeling a wide range of devices.
While gray-box models again achieve results that are not on par with black-box ones, therefore requiring further exploration.

\subsection{Efficiency}
\label{sec:efficiency}

\begin{figure*}[t]
    \begin{minipage}[b]{.33\textwidth}
        \centering
        \includegraphics[width=\textwidth]{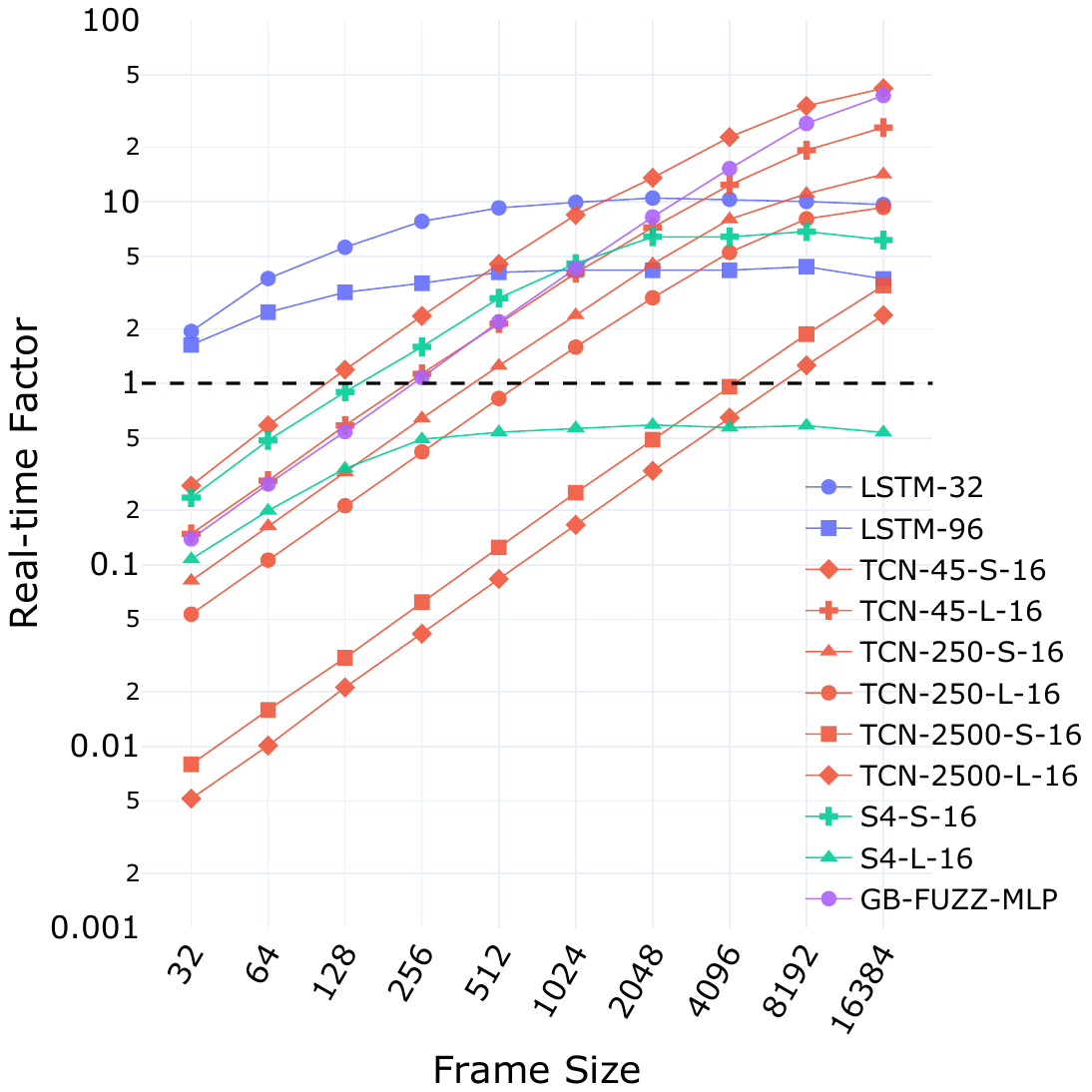}
        \\\textbf{(a)}
    \end{minipage}
    \begin{minipage}[b]{.33\textwidth}
        \centering
        \includegraphics[width=\textwidth]{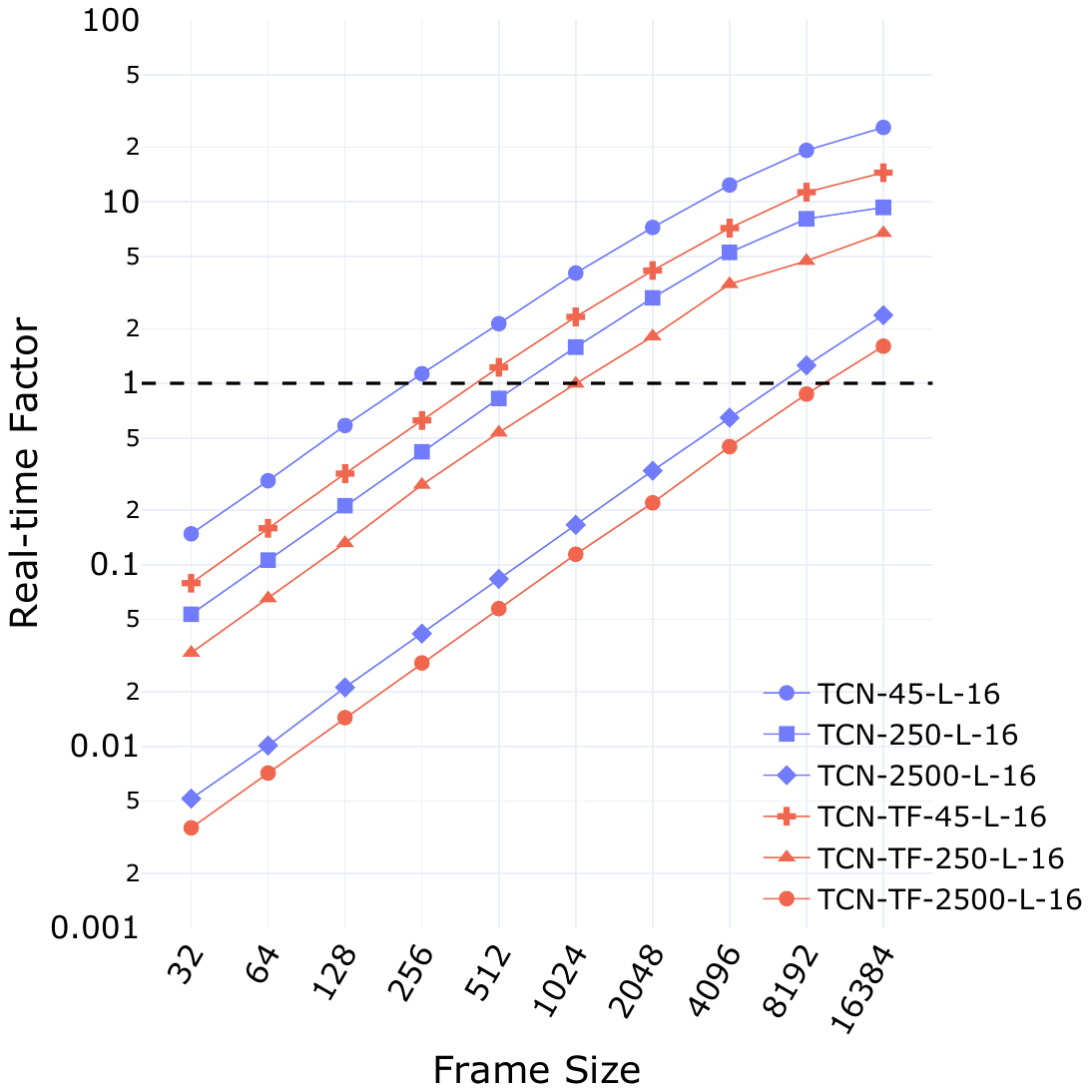}
        \\\textbf{(b)}
    \end{minipage}
    \begin{minipage}[b]{.33\textwidth}
        \centering
        \includegraphics[width=\textwidth]{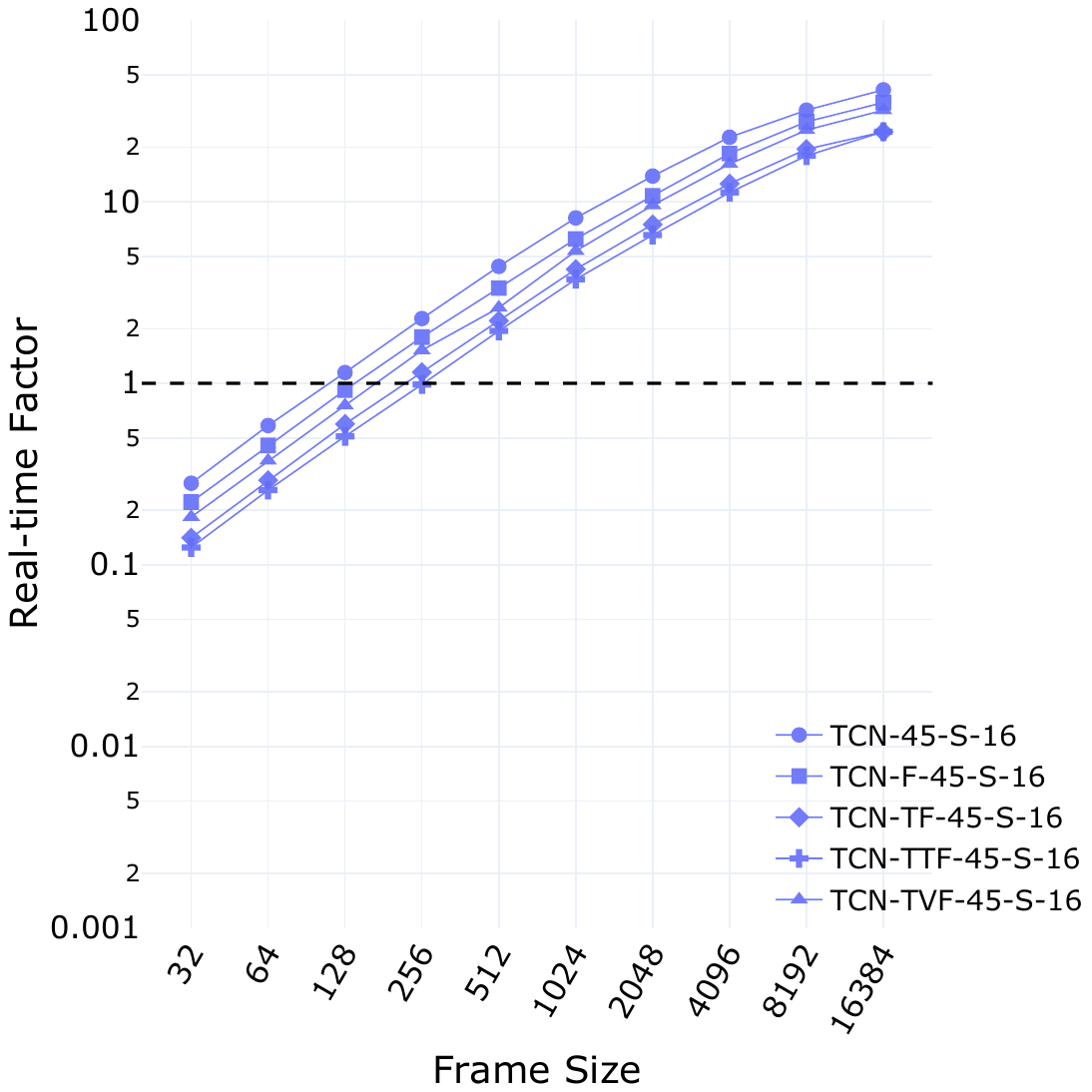}
        \\\textbf{(c)}
    \end{minipage}
    \caption{Real-time factor vs. Frame size for some of the models included in the experiments}
    \label{fig:rtf_vs_frame}
\end{figure*}

We investigated the run-time of the proposed models and conditioning methods in a frame-based implementation to mimic a standard digital audio effect implementation. 
The real-time factor (RTF) is defined as:
\begin{equation}
    \text{RTF}:=\frac{F}{T \cdot f_{s}}
\end{equation}
where $F$ is the frame size in samples and $T$ is the time in seconds to process them at sampling rate $f_{s}$. 
We measure the RTF at power of 2 frame sizes, $F \in 32, 64, ..., 16384$, on a 2019 MacBook Pro with an Intel Core i5 @ 1.4 GHz.

Results for non-parametric models are shown in Fig.~\ref{fig:rtf_vs_frame}a and \ref{fig:rtf_vs_frame}b and for parametric ones in Fig.~\ref{fig:rtf_vs_frame}c.
We show only a limited number of configurations for clarity.
In a block-based formulation, TCN models require a buffer of past samples such that we pass an input of $F + r - 1$ samples, where $F$ is the number of output samples and $r$ is the receptive field in samples.

LSTMs achieve real-time operation at every frame size, but the RTF remains constant over a certain size due to the inability to parallelize computations along the temporal dimension.
The RTF for TCN, GCN (not shown) and GB models, is instead proportional to the frame size.
For convolutional backbones, whether the models achieve real-time operation at a sufficiently small frame size is strongly related to the receptive field and the number of blocks.
Interestingly, S4 models show a behavior in between recurrent and convolutional networks, where the RTF is proportional to frame size for small sizes and flattens out for large sizes.
Although best performing, S4 models require appropriate choice of number of blocks and state dimension to allow real-time operation, in fact, our larger model does not achieve $\text{RTF}> 1$.
In our implementation, all proposed gray-box models (of which we only show GB-FUZZ-MLP) show very similar RTF, proportional to frame size.

Fig.~\ref{fig:rtf_vs_frame}b compares TCN models with and without TFiLM conditioning, with temporal feature-wise modulation consistently reducing the RTF at every model and frame size.
Also, in Fig.~\ref{fig:rtf_vs_frame}c we compare different conditioning methods (FiLM, TFiLM, TTFiLM, TVFiLM) for parametric TCN models, and include a non-parametric model (TCN-45-S-16) for reference.
We notice how, even if computationally as efficient as models with FiLM and TVFiLM, TTFiLM conditioning is the worst in terms of RTF, while TVFiLM achieves real-time performance between FiLM and TFiLM, introducing time-varying conditioning at a low computational cost. 

Our results represent a worse-case scenario, since optimized C++ implementations may achieve a speedup compared to the PyTorch models used in our analysis \citep{wright2019real}.

\subsection{Subjective Evaluation}
\label{sec:subj_eval}

\begin{figure*}[t]
    \begin{minipage}{1\textwidth}
        \centering
        \includegraphics[height=5cm]{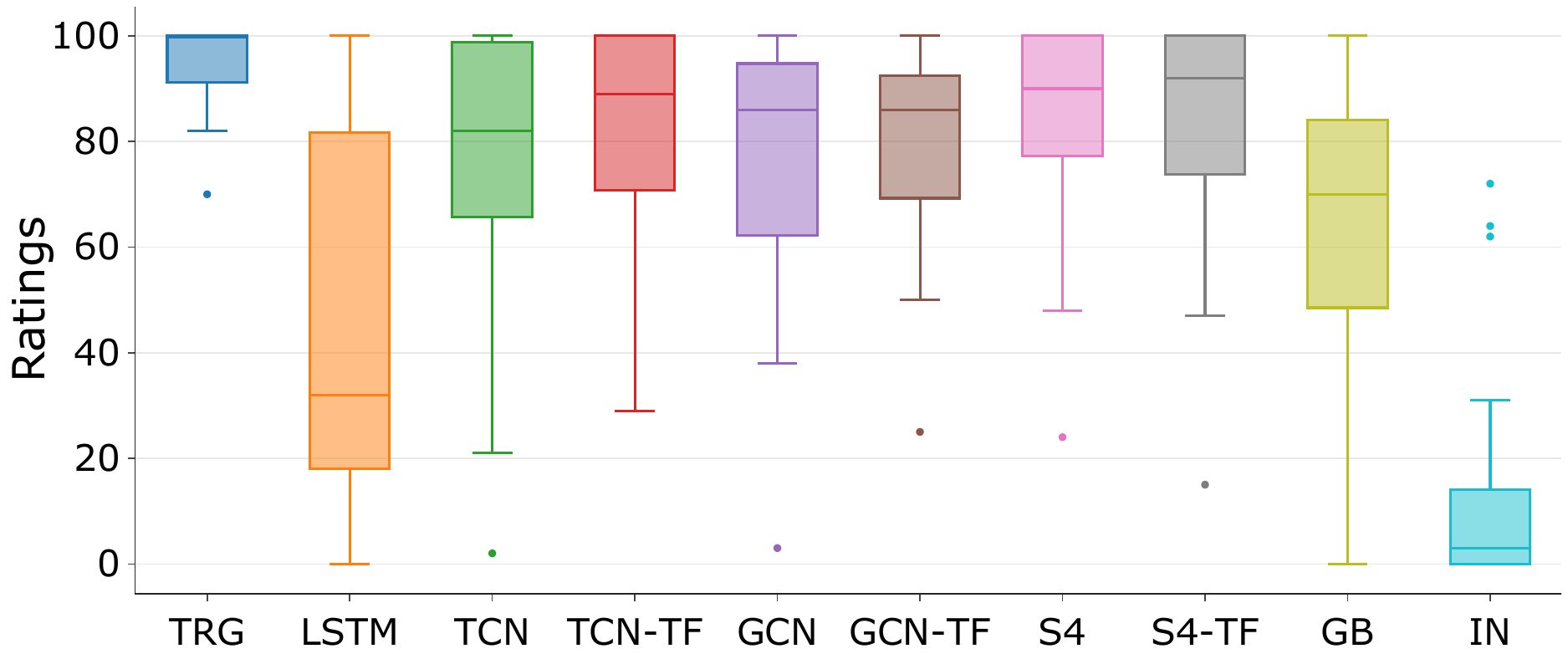}
        \\\textbf{(a)} Overall
    \end{minipage}\\\\
    
    \begin{minipage}{.255\textwidth}
        \centering
        \includegraphics[height=4.1cm]{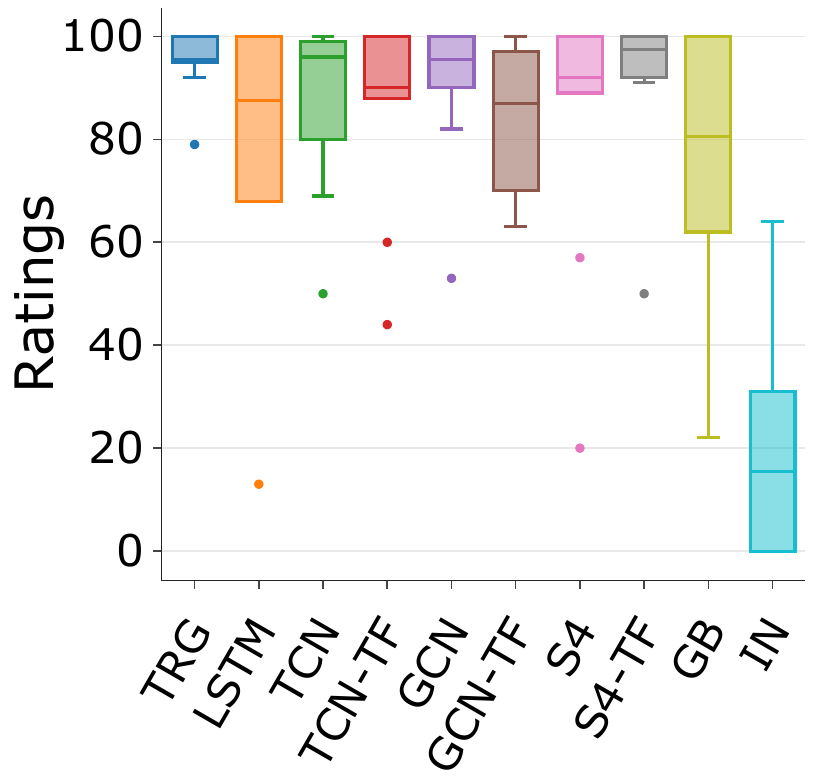}
        \\\textbf{(b)} Compressor/Limiter
    \end{minipage}
    \begin{minipage}{.23\textwidth}
        \centering
        \includegraphics[height=4.1cm, trim=31 0 0 0, clip]{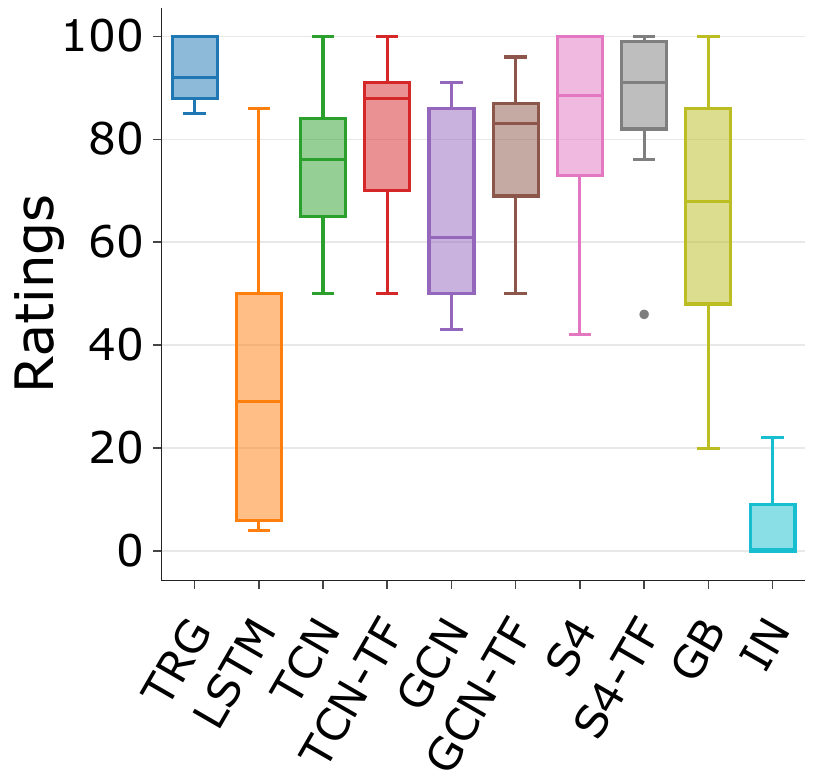}
        \\\textbf{(c)} Overdrive
    \end{minipage}
    \begin{minipage}{.23\textwidth}
        \centering
        \includegraphics[height=4.1cm, trim=31 0 0 0, clip]{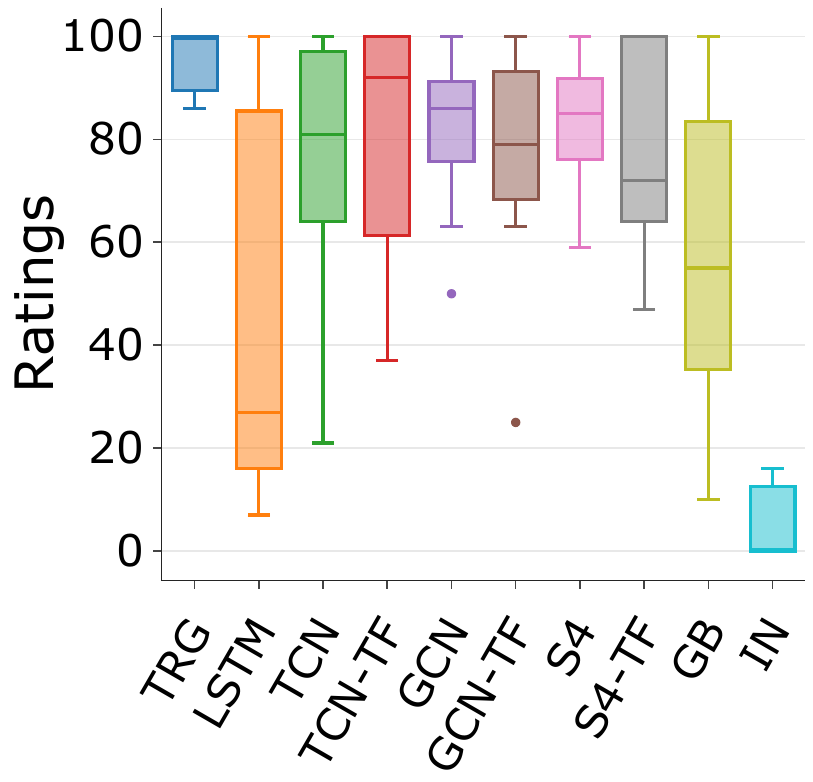}
        \\\textbf{(d)} Distortion
    \end{minipage}
    \begin{minipage}{.23\textwidth}
        \centering
        \includegraphics[height=4.1cm, trim=31 0 0 0, clip]{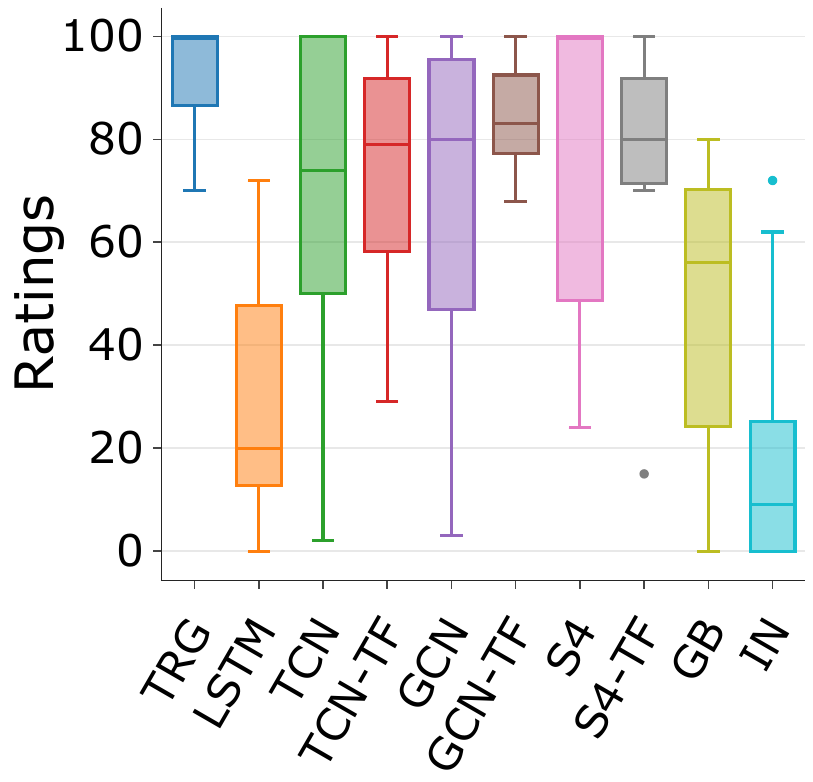}
        \\\textbf{(e)} Fuzz
    \end{minipage}
    \caption{Subjective ratings for different architectures: overall and for each effect type}
    \label{fig:list_test}
\end{figure*}

To evaluate the perceptual differences among architectures, we conducted a MUSHRA listening test. 
Each trial presented participants with processed audio from all architectures under evaluation (LSTM, TCN, TCN-TF, GCN, GCN-TF, S4, and S4-TF) alongside a reference signal (wet-output) and an anchor (dry-input). 
We selected 12 effects (3 per type) out of the 16 we trained on.
For each effect, we curated three distinct sound examples designed to highlight model performance. 
These included a bass riff with a high input level and clear attacks to emphasize timbral and temporal characteristics, a guitar sample at similarly high input levels with pronounced attacks to further assess these aspects, and a second guitar sample with a very low input level and steady picking. 
This last example was chosen to evaluate the models’ ability to faithfully capture devices across extreme input dynamics. 

For each architecture we selected the best-performing model in terms of total test loss.
The listening test consisted of a total of 68 possible questions, with each participant randomly assigned seven to maintain a manageable test duration. 
A total of 27 participants were recruited, and a rigorous screening process was implemented. 
Participants were assessed on prior experience with listening tests, musical training or production experience, and familiarity with the specific effects being evaluated (compressors, overdrive/distortion, and fuzz). 
Further filtering was applied to ensure test reliability, participants who did not rate the reference sample at 80 or higher in more than one out of their seven trials were excluded from the final analysis. After applying these screening criteria, we obtained 350 valid ratings in total, averaging 35 per architecture and approximately 8.75 ratings per effect type.
Our strict participant selection strategy was intentional, prioritizing fewer but highly skilled listeners. 
This approach significantly reduced rating variance, leading to a more reliable subjective evaluation of model performance.

We show the overall results in Fig.~\ref{fig:list_test} together with ratings broken down by effect category.
In general, subjective ratings confirm the results obtained during the training process, with S4 architectures performing best, followed by convolutional backbones, gray-box models and LSTMs.
TFiLM conditioning allows to increase median accuracy and lower variance, improving modeling reliability across effects types and devices.
With respect to objective metrics, gray-box models are on average rated higher than LSTMs, but with high variance, which might entail the potential to match black-box approaches while requiring careful architecture design for different effect types.
Once again, LSTMs are confirmed to be unreliable as modeling architectures on a wide range of devices.

If we break down ratings by effect type we notice that in the majority of cases TFiLM helps to improve performance, in particular for convolutional backbones trained on overdrive, distortion and fuzz.
Proposed gray-box models perform, on average, well on compression and fairly on overdrive, while less so on distortion and fuzz.
Also, GCNs are not consistently better than TCNs, which confirms that the added complexity of GCNs is in general not necessary for good performance.
S4 models remain the most consistent, with median ratings at or above 80 in all but one case.

\subsection{Correlation}
\label{sc:correlation}

We further extend our analysis calculating the Spearman correlation between objective metrics and subjective ratings (Table~\ref{tab:corr_metrics-fad-ratings}).
Out of the signal-based metrics, MR-STFT shows the highest correlation with human ratings, supporting the assumption that it is a suitable term to include in the loss function.
MAPE and L1 are the second and third most correlated metrics, again supporting the idea that including a time-domain loss term is meaningful for modeling tasks.
Considering effects type, MR-STFT is the most relevant for high-distortion effects, while MAPE and L1 are the most correlated with subjective ratings for overdrive and compression respectively, although in the latter case, absolute correlation values are very low or not significant ($p > 0.05$), showing no single metric to be well suited for training.
We conclude that using frequency-domain in conjunction with time-domain losses seems to be a sensible choice for a wide variety of cases, but that improvements are possible, especially for compression modeling.

Observing latent representation-based metrics instead, VGGish and AFx-Rep vectors are overall the most correlated with subjective ratings.
Looking at specific effect types, none of the representations seem to be suited for compression and overdrive models evaluation, while CLAP and VGGish are, respectively, highly correlated with distortion and fuzz effects.

\setlength{\tabcolsep}{3pt}
\begin{table*}[t]
    \small
    \caption{Spearman correlation ($\rho$) and $p$-value ($<0.05$) between subjective ratings and objective metrics for each device type and overall. Correlation is negative since low metrics values correspond to high ratings values. Bold indicates highest absolute value.}
    \label{tab:corr_metrics-fad-ratings}
    \centerline{
        \begin{tabular}{l cc cc cc cc cc cc cc cc cc} 
            \midrule
            \midrule
            \multirow{2}{*}{Effect Type} 
            & \multicolumn{2}{c}{L1} 
            &  \multicolumn{2}{c}{MR-STFT} 
            & \multicolumn{2}{c}{MSE} 
            & \multicolumn{2}{c}{ESR} 
            & \multicolumn{2}{c}{MAPE}
            & \multicolumn{2}{c}{FAD$_{\text{VGGish}}$} 
            &  \multicolumn{2}{c}{FAD$_{\text{PANN}}$} 
            & \multicolumn{2}{c}{FAD$_{\text{CLAP}}$} 
            & \multicolumn{2}{c}{FAD$_{\text{AFx-Rep}}$} \\
            
            \cmidrule(lr){2-3} 
            \cmidrule(lr){4-5} 
            \cmidrule(lr){6-7} 
            \cmidrule(lr){8-9}
            \cmidrule(lr){10-11}
            \cmidrule(lr){12-13}
            \cmidrule(lr){14-15}
            \cmidrule(lr){16-17}
            \cmidrule(lr){18-19}
            
            & $\rho$ & $p$ 
            & $\rho$ & $p$ 
            & $\rho$ & $p$ 
            & $\rho$ & $p$ 
            & $\rho$ & $p$
            & $\rho$ & $p$
            & $\rho$ & $p$
            & $\rho$ & $p$
            & $\rho$ & $p$ \\
            
            \hline
            Comp./Limit. 
                & \vline~\textbf{-0.188} & \ding{51} 
                & -0.158 & \ding{51} 
                & -0.185 & \ding{51} 
                & -0.149 & \ding{55} 
                & -0.152 & \ding{55}
                & \vline~-0.188 & \ding{51} 
                & -0.193 & \ding{51} 
                & -0.172 & \ding{51} 
                & \textbf{-0.194} & \ding{51} \\
            Overdrive 
                & \vline~-0.183 & \ding{51} 
                & -0.277 & \ding{51} 
                & -0.171 & \ding{51} 
                & -0.264 & \ding{51} 
                & \textbf{-0.361} & \ding{51}
                & \vline~-0.207 & \ding{51} 
                & -0.061 & \ding{55} 
                & -0.182 & \ding{51} 
                & \textbf{-0.211} & \ding{51} \\
            Distortion 
                & \vline~-0.242 & \ding{51} 
                & \textbf{-0.536} & \ding{51} 
                & -0.267 & \ding{51} 
                & -0.389 & \ding{51} 
                & -0.214 & \ding{51}
                & \vline~-0.592 & \ding{51} 
                & -0.404 & \ding{51} 
                & \textbf{-0.613} & \ding{51} 
                & -0.533 & \ding{51} \\
            Fuzz 
                & \vline~-0.424 & \ding{51} 
                & \textbf{-0.488} & \ding{51} 
                & -0.406 & \ding{51} 
                & -0.470 & \ding{51} 
                & -0.453 & \ding{51}
                & \vline~\textbf{-0.511} & \ding{51} 
                & -0.314 & \ding{51} 
                & -0.434 & \ding{51} 
                & -0.482 & \ding{51} \\
            \hline
            Overall 
                & \vline~-0.332 & \ding{51} 
                & \textbf{-0.427} & \ding{51} 
                & -0.314 & \ding{51} 
                & -0.311 & \ding{51} 
                & -0.339 & \ding{51}
                & \vline~\textbf{-0.430} & \ding{51} 
                & -0.303 & \ding{51} 
                & -0.406 & \ding{51} 
                & -0.417 & \ding{51} \\
            \hline
            \hline
        \end{tabular}
    }
\end{table*}


\section{Conclusions and Future Work}
\label{sec:conclusion}
In this work, we conducted extensive experiments to establish the state of the art in differentiable black-box and gray-box approaches for nonlinear audio effects modeling.
Our study explored task-specific neural network architectures, proposed differentiable DSP-based gray-box models, and presented ToneTwist AFx, a large dataset of dry-input/wet-output pairs for audio effects research.

We described the challenges of training and comparing different approaches, and demonstrated how careful consideration of the training scheme is essential to ensure fair comparisons and draw reliable conclusions.\\
Thanks to our extensive objective and subjective evaluations we concluded that state-space models are the most suited to model a wide range of devices, with TCN and GCN architectures also achieving very good results.
We found LSTM-based recurrent networks to be unreliable across devices and DDSP-based models not yet on par with neural networks.
Time-varying conditioning like TFiLM was proven beneficial, across convolutional and state-space backbones, to improve median and variance in objective metrics and subjective ratings.
Our analysis of the correlation between subjective ratings and signal-based metrics showed that a loss function combining time-domain (L1/MAPE) and spectral-domain (MR-STFT) is generally effective. 
However, there is room for improvement in signal-based and latent representation-based metrics for model evaluation.

Given these results, future work might focus on any of the following aspect:
\begin{itemize}
    \item Identify perceptually relevant signal-based and latent representation-based metrics for audio effects modeling. This would enable optimized training schemes for each architecture while ensuring objective model comparisons.
    \item Enhance gray-box models to achieve modeling accuracy on par with black-box approaches.
    \item Further study parametric modeling and conditioning methods across a wide range of effects, especially focusing on efficient implementations.
    \item Improve the implementation of gray-box processors and controllers to speed-up processing and increase overall efficiency.
    \item Investigate universal audio effects modeling architectures that can encompass all effect categories, including linear, nonlinear, and modulation effects.
    \item Integrate hyperparameter optimization, along with model pruning or distillation, to achieve high-performance and efficient models.
\end{itemize}

\section{Acknowledgments}
Funded by UKRI and EPSRC as part of the “UKRI CDT in Artificial Intelligence and Music”, under grant EP/S022694/1.

\bibliographystyle{unsrtnat}
\bibliography{references} 


\end{document}